\documentclass[journal]{IEEEtran}

\usepackage{booktabs}
\usepackage{amsmath}
\usepackage{amsthm}
\usepackage{multirow}
\usepackage{color,cite}
\usepackage{subfigure}
\usepackage{mathrsfs}
\usepackage{amssymb}
\usepackage{bm}
\usepackage{graphicx}
\usepackage{amsfonts}
\usepackage{algorithm}
\usepackage{algorithmic}
\usepackage{epstopdf}
\usepackage{geometry}
\usepackage{hyperref}
\newtheorem{theorem}{Theorem}
\newtheorem{corollary}{Corollary}

\begin{document}

\bibliographystyle{IEEEtran}

\title{\huge Near-Field Channel Estimation in Dual-Band XL-MIMO with Side Information-Assisted Compressed Sensing}
\author{
    Haochen~Wu, Liyang~Lu,~\IEEEmembership{Member,~IEEE}, and Zhaocheng~Wang,~\IEEEmembership{Fellow,~IEEE} \\
    \thanks{This work was supported in part by the Postdoctoral Fellowship Program of CPSF under Grant Number GZC20231374, in part by the China Postdoctoral Science Foundation under Grant Number 2024M751680. (\emph{Corresponding author: Zhaocheng Wang and Liyang Lu})}
    \thanks{H. Wu and L. Lu are with the Department of Electronic Engineering, Tsinghua University, Beijing 100084, China (e-mail: wuhc23@mails.tsinghua.edu.cn, luliyang@mail.tsinghua.edu.cn).}
    \thanks{Z. Wang is with the Department of Electronic Engineering, Tsinghua University, Beijing 100084, China, and also with the Shenzhen International Graduate School, Tsinghua University, Shenzhen 518055, China (e-mail:zcwang@tsinghua.edu.cn).}
    \thanks{H. Wu and L. Lu contributed equally to the project and should be considered co-first authors.}
}
\maketitle

\begin{abstract} \label{Abstract}
    Near-field communication comes to be an indispensable part of the future sixth generation (6G) communications at the arrival of the forth-coming deployment of extremely large-scale multiple-input-multiple-output (XL-MIMO) systems.
    Due to the huge array aperture and high-frequency bands, the electromagnetic radiation field is modeled by the spherical waves instead of the conventional planar waves,
    leading to severe weak sparsity to angular-domain near-field channel.
    Therefore, the channel estimation reminiscent of the conventional compression sensing (CS) approaches in the angular domain,
    judiciously utilized for low pilot overhead, may result in unprecedented challenges.
    To this end, this paper proposes a brand-new near-field channel estimation scheme by exploiting the naturally occurring useful side information.
    Specifically, we formulate the dual-band near-field communication model based on the fact that high-frequency systems are likely to be deployed with lower-frequency systems.
    Representative side information, i.e., the structural characteristic information derived by the sparsity ambiguity and
    the out-of-band spatial information stemming from the lower-frequency channel,
    is explored and tailored to materialize exceptional near-field channel estimation.
    Furthermore, in-depth theoretical analyses are developed to guarantee the minimum estimation error,
    based on which a suite of algorithms leveraging the elaborating side information are proposed.
    Numerical simulations demonstrate that the designed algorithms provide more assured results than the off-the-shelf approaches in the context of the dual-band near-field communications
    in both on- and off-grid scenarios,
    where the angle of departures/arrivals are discretely or continuously distributed, respectively.
\end{abstract}


\IEEEpeerreviewmaketitle

\section{Introduction} \label{S1}

\IEEEPARstart{E}{xtremely} large-scale multiple-input-multiple-output (XL-MIMO) is expected to fulfill the demands for ubiquitous connecting of the sixth generation (6G) mobile networks \cite{wangzhe2024survey}.
Due to the large spatial multiplexing gain offered by the exploitation of large number of antennas, which is much more than that of massive MIMO in the current fifth-generation (5G) communications,
XL-MIMO can provide a 10-fold increase in spectral efficiency \cite{dai2022tcom}.
Additionally, wideband communications are prospective for more available bandwidth owing to rich spectrum resources at high-frequency bands, e.g.,
millimeter-wave (mmWave) and terahertz (THz) bands \cite{ziyuansha2021}.
In conjunction with the flexible deployment of high-frequency antennas,
wideband XL-MIMO is capable of providing the aforementioned benefits,
which is regarded as an essential component in future 6G mobile networks \cite{dai2023mag}.

Due to the extremely large array aperture and high-frequency bands,
the electromagnetic (EM) characteristics of wideband XL-MIMO undergo a fundamental change \cite{rayleighdistance},
where the EM radiation field can be partitioned into the far-field and near-field regions.
The boundary between these two regions is approximately determined by the Rayleigh distance which is proportional to the product of the square of array aperture and carrier frequency \cite{rayleighdistance,dai2023mag}.
Outside the Rayleigh distance is the far-field region, where the EM waves can be modeled by the planar waves, e.g., the wave model used in massive MIMO systems \cite{swang2022jsac}.
Within the Rayleigh distance, near-field propagation becomes dominant, hence the EM waves need to be accurately modeled by spherical waves \cite{dai2023mag}.

However, huge numbers of antennas cause substantial pilot overhead in channel estimation.
To address this issue, the compressed sensing (CS) technique is capable of providing accurate channel estimation while maintaining low pilot overhead by fully exploiting the sparsity of the channel.
In the conventional massive MIMO systems considering planar wave model,
the number of significant paths is much smaller than the number of antennas, hence the channel exhibits sparsity in the angular domain \cite{zhao2016access}.
Nevertheless, for the near-field region, spherical waves introduce a distance ingredient, causing severe sparsity ambiguity in angular-domain representation,
which indicates that one single near-field path component spreads towards multiple angles \cite{dai2022tcom}.
Then, the near-field channel in the angular domain exhibits weak sparsity, i.e., the number of nonzero channel-taps becomes relatively large.
It precludes accurate channel estimation in the angular domain through the conventional CS,
since the ratio of the nonzero channel-taps in the channel vector is greater than the upper bound of sparsity required for accurate estimation \cite{liyang2022}.

Many research efforts have been devoted to mitigating the aforementioned weak sparsity of the near-field channel \cite{dai2022tcom,chenyuanbin2023,ylu2023,xwei2022}.
For instance, a polar-domain transformation is proposed in \cite{dai2022tcom} by sampling the angle uniformly and sampling the distance non-uniformly.
The near-field channel is then mapped into the polar domain, involving both angular and distance ingredients,
which is sparse enough for accurate estimation by the conventional CS.
In \cite{chenyuanbin2023}, the authors propose to decompose the near-field channel into triple parametric variants firstly,
and then use the CS algorithm to estimate the triple parameters of the channel.
The works \cite{ylu2023} and \cite{xwei2022} continue to exploit the polar-domain transformation \cite{dai2022tcom} for efficient channel estimation in the non-line-of-sight (NLoS) and the mixed line-of-sight (LoS) and NLoS environments, respectively.
Despite of these achievements providing sufficiently sparse signals that can be recovered by CS algorithms,
the bottleneck is that the dimensions of the intrinsic signal processing problems are significantly increased.
Quantitatively, when the number of antennas at the base station (BS) is $256$, the number of the polar-domain channel-taps is about $2200$,
which is $4$ times more than the number of conventional angular-domain channel-taps based on Discrete Fourier Transformation (DFT) \cite{dai2022tcom}.
The excessively high dimensions of signal processing problems cause unacceptable complexity,
which precludes these approaches from alleviating weak sparsity in practice.

Actually, useful information stemming from the signal itself or communication environments, called side information \cite{Wyner1976},
can help mitigate the weak sparsity of angular-domain near-field channels without increasing dimensions in the signal processing problems.
The side information can provide the emphatic correlation between the nonzero channel-taps and the pilot training matrix,
and also the confirmed potential to filter out noise interference, which significantly improves the sparsity bound required for accurate estimation,
leading to more desirable performance compared with the traditional methods without side information.

There are two types of key side information for near-field channel estimation in terms of the channel-tap characteristics and the communication system architecture.
On one hand, angular-domain sparsity ambiguity offers continuous nonzero channel-taps, which is considered as the structural characteristic information, i.e., block structure \cite{liyang2022}.
Meanwhile, for multi-carrier communication systems, the positions of nonzero channel-taps are typically the same \cite{dai2022tcom},
which indicates that the near-field channel matrix always exhibits block sparsity with same sparse patterns among different subcarriers.
Current researches have proved that the exploitation of block structure provides more reliable recovery performance than the conventional non-block approaches,
since it improves the upper bound of the sparsity level required for reliable recovery \cite{liyang2023arxiv}.
On the other hand, high-frequency systems are likely to be deployed at the same location as lower-frequency systems,
giving rise to out-of-band spatial information \cite{outofband2017,outofband2018}.
A representative example is that millimeter wave (mmWave) systems are typically deployed together with Sub-6GHz systems for providing wide area control signals and multi-band communications \cite{outofband2018}.
Then, the out-of-band spatial information in lower-frequency Sub-6GHz systems is useful because the spatial characteristics of mmWave and Sub-6GHz channels are similar \cite{outofbandpdfPeter2016}.
Specifically, this out-of-band spatial information can be formulated as the weights for index selection of nonzero channel-taps in weighted channel estimation \cite{Scarletttsp2013}.

Nevertheless, the investigation of low-complexity near-field channel estimation assisted by side information is still in its infancy.
Firstly, to the best of our knowledge, there is no system modeling of near-field communications adopting out-of-band spatial information.
As the near-field region of lower-frequency communication systems is smaller than that of high-frequency systems,
the user may stay in either the near-field or far-field region of the lower-frequency system,
and always in the near-field region of the high-frequency system.
In this dual-band communication system,
the out-of-band spatial information may come from either near-field or far-field channels of the lower-frequency system.
The formulation of this hybrid communication model, and the extraction and exploitation of the out-of-band spatial information remain to be solved.
Secondly, the methodology, which employs both out-of-band spatial and structural characteristic information, needs to be carefully designed.
Current works using the side information for weighted sparse recovery, e.g., \cite{Scarletttsp2013,outofband2018,liyang2019wcnc,liyang2019},
do not consider the complex signals in practical communication scenarios,
wherein \cite{Scarletttsp2013,outofband2018} do not even take block structure into account.
It is evident that these existing studies may not be suitable for the out-of-band spatial information- and block structure-assisted near-field channel estimation.

Against the above backgrounds, this paper proposes the scheme of side information-assisted dual-band near-field channel estimation in the angular domain.
The main contributions are summarized as follows.

\begin{enumerate}
    \item The model of dual-band near-field communication systems is formulated.
          Specifically, the scenario where lower-frequency and high-frequency systems are deployed simultaneously at the same location is highlighted,
          where the various frequencies result in different near-field regions.
          Hence, two communication scenarios, i.e., near-field channel estimation aided by far-field and near-field out-of-band spatial information respectively, are further discussed.
          The elaborating dual-band model paves the way for the subsequent analysis and estimation procedures.

    \item Theoretical analysis of reliable estimation assisted by side information is developed.
          We begin with embedding the dedicated side information and the practical complex channel consideration into the iterative mechanism of the orthogonal matching pursuit (OMP) algorithm,
          leading to a prior factor for correct index selection of nonzero channel-taps.
          The analysis of minimizing the estimation error corresponding to the prior factor is derived,
          which involves judiciously high-dimensional $\chi^2$ approximation for tighter theoretical guarantees.

    \item A brand-new near-field channel estimation scheme is proposed,
          wherein a series of algorithms derived from the OMP
          leveraging the naturally occurring side information, are developed.
          Owing to the exploitation of the prior factor derived,
          the index selections of nonzero channel-taps of the designed algorithms are more accurate,
          leading to a higher upper bound of the sparsity level required for reliable recovery.
          This results in more desirable estimation performance than the approaches without the assistance of side information
          in both on- and off-grid scenarios, where the angles of arrivals/departures are discretely or continuously distributed correspondingly.
\end{enumerate}

The rest of the paper is organized as follows.
Section~\ref{S2} introduces the system model and problem formulation, followed by some useful definitions.
In Section~\ref{S3}, we start with the complex logit-weighted OMP (CLW-OMP)
and propose our complex simultaneous logit-weighted block OMP (CSLW-BOMP) algorithm.
Moreover, the performance guarantees for these algorithms are derived.
In Section~\ref{S4}, the simulation results are presented and analyzed,
and the conclusions are drawn in Section~\ref{S5}.

\emph{Notation}: We briefly summarize the notations used in this paper.
Boldface lowercase letters, e.g., $\mathbf{x}$, denote vectors,
and boldface uppercase letters denote matrices, e.g., $\mathbf{X}$.
Calligraphic letters, e.g., $\mathcal{S}$, are used for sets,
and non-boldface letters e.g., $x$ and $X$, represent scalars.
Moreover, $\mathbf{X}_{\mathcal{S}}$ represents the submatrix of $\mathbf{X}$ composed of the column vectors whose indices are from $\mathcal{S}$.
Superscript $\rm H$ denotes the conjunction transpose,
and $\mathbf{0}$ represents the all zero vector or matrix.
$\mathcal{CN}(a,b)$ denotes a complex Gaussian distribution with mean $a$ and variance $b$,
$\chi^2(k)$ denotes a central $\chi^2$ distribution with $k$ degrees of freedom,
$\chi'^2(k,\lambda)$ denotes the non-central $\chi^2$ distribution with its degree of freedom being $k$ and noncentrality parameter being $\lambda$,
and $[\mathcal{CN}(a_1,b_1),\mathcal{CN}(a_2,b_2),\cdots,\mathcal{CN}(a_n,b_n)]$ denotes a vector with length $n$,
where the $i$-th entry follows $\mathcal{CN}(a_i,b_i)$.
We use $\|\cdot\|_F$, $|\cdot|$, $\arg(\cdot)$ to denote the Frobenius norm, modulus and argument of their objective, respectively.
The variables from the Sub-6GHz band are underlined, e.g., $\underline{\mathbf{X}}$, for clarity.

\section{System Model and Mathematical Formulation} \label{S2}

In this section, we first provide the dual-band XL-MIMO system model,
and then introduce the corresponding mathematical formulation.

\begin{figure}[bp!]
    \begin{center}
        \includegraphics[width=0.5\textwidth]{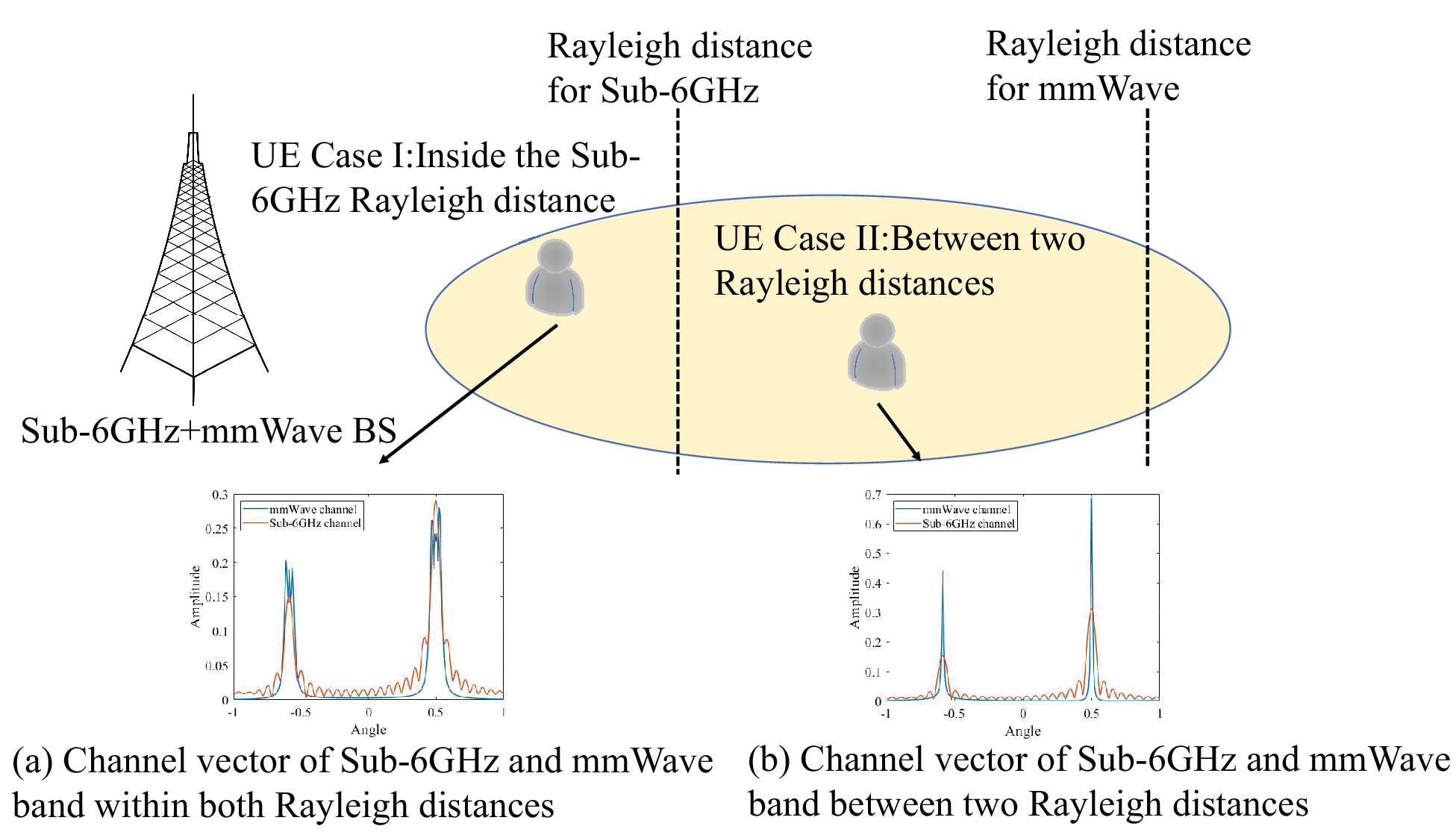}
    \end{center}
    \caption{System model and schematic.}
    \label{Model}
\end{figure}

\subsection{System Model}
\label{S21}
As depicted in Fig.~\ref{Model}, we formulate the system model of dual-band XL-MIMO deployed in the time division multiplexing (TDD) mode,
where mmWave and Sub-6GHz bands are exemplified as the different frequency band solutions.
The antennas on the mmWave and Sub-6GHz bands are set to be co-located, aligned and have same apertures \cite{outofband2018}.
The BS is composed of $N$ antennas on the mmWave band and $\underline{N}$ antennas on the Sub-6GHz band,
while each UE has only one antenna.
In order to ensure same antenna apertures for different frequency bands, $N$ should be larger than $\underline{N}$.
Specifically, the BS and user equippment (UE) are deployed both on mmWave and Sub-6GHz bands with $K$ subcarriers,
whose wavelengths are denoted as $\underline{\lambda}$ and $\lambda$, respectively.
Due to frequency differences, $\underline{\lambda}>\lambda$.
The antenna spacing of BS is $\lambda/2$ on the mmWave band and $\underline{\lambda}/2$ on the Sub-6GHz band.
As the boundary of the far-field and the near-field regions, i.e., the Rayleigh distance $Z=\frac{2D^2}{\lambda}$,
is proportional to the carrier frequency and array aperture $D$ \cite{rayleighdistance},
the Rayleigh distance of Sub-6GHz systems is smaller than that of the mmWave systems due to same antenna apertures and larger carrier wavelengths.
Moreover, both frequency bands have considerably large Rayleigh distances,
leading to different channel models presented as follows.

In conventional far-field scenarios,
the electromagnetic radiaition field is modeled by the planar waves.
Assume that there exist $L$ significant paths, the far-field channel $\mathbf{h}_{far} \in \mathbb{C}^{N \times 1}$ can be given by \cite{Angular1,Angular2}
\begin{equation}
    \mathbf{h}_{far}=\sqrt{\frac{N}{L}}\sum_{l=1}^Lg_le^{-jk_mr_l}\mathbf{a}(\theta_l).
    \label{Planar}
\end{equation}
Since the number of antennas is much larger than that of the significant paths,
the far-field channel exhibits sparsity in the angular domain,
i.e., the number of nonzero channel-taps is much smaller than that of the total channel-taps \cite{Sparse}.
Nevertheless, due to the large antenna aperture $D$ and the high-frequency carrier in XL-MIMO systems,
the Rayleigh distance is significantly increased \cite{dai2022tcom}.
For example, when the carrier frequency is 100GHz and the antenna aperture is 0.5m,
the Rayleigh is about 167m, which can cover a whole cell.
Consequently, the UE is more likely to fall within the near-field regions
where the spherical wave model should be adopted to accurately represent the electromagnetic propagation characteristics.
Hence the near-field channel vector $\mathbf{h}_k \in \mathbb{C}^{N \times 1}$ can be written as \cite{Spherical}
\begin{equation}
    \mathbf{h}_k=\sqrt{\frac{N}{L}}\sum_{l=1}^{L}g_le^{-jk_mr_l}\mathbf{b}(\theta_l,r_l).
    \label{Spherical}
\end{equation}

\begin{figure}[tp!]
    \begin{center}
        \includegraphics[width=0.5\textwidth]{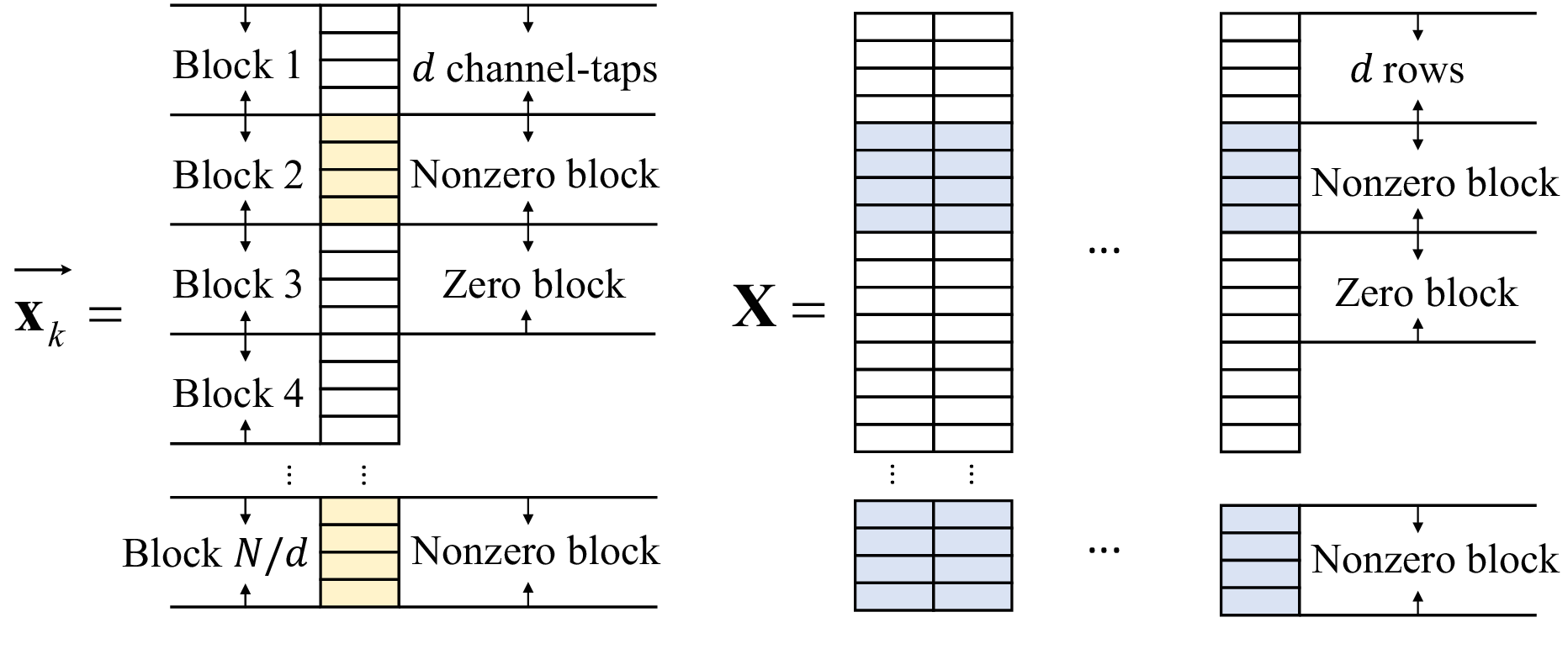}
    \end{center}
    \caption{Schematic diagram of block sparse vector $\mathbf{x}_k$ and matrix $\mathbf{X}$ with BMMV model.}
    \label{Sparsity}
\end{figure}

Different from the model in (\ref{Planar}), the near-field steering vector $\mathbf{b}(\theta_l,r_l)$ depends not only on the angle $\theta_l$ but also on the radial distance $r_l$.
Under such circumstance, $\mathbf{b}(\theta_l,r_l)$ exhibits nonlinearity with respect to the antenna index $l$,
and hence should be represented by multiple far-field Fourier vectors instead of a single one.
Consequently, the nonzero channel-taps will spread towards multiple angles and exhibits weak sparsity,
where the ratio of nonzero channel-taps is relatively large and exceeds the upper bound of the reconstructible sparsity required for reliable recovery of the conventional CS algorithms \cite{dai2022tcom}.
Fortunately, this weak sparsity in angular-domain near-field channels can be regarded as the block-sparse structure,
which can be leveraged to provide more reliable estimation performance.
As illustrated in Figs.~\ref{Model}(a) and \ref{Model}(b), the nonzero channel-taps of near-field channels naturally occur in blocks,
which is formulated as $\mathbf{x}_k$ in the left-hand side of Fig.~\ref{Sparsity} \cite{Block1,Block2}.

Therefore, the different positions of the UE to the BS induce two distinct cases of the dual-band XL-MIMO system,
which are presented as follows.

\textbf{\textit{Case I}}:
The UE is situated within the Rayleigh distance of both the mmWave and Sub-6GHz bands.
Under such circumstance, since both the mmWave and Sub-6GHz channels experience the near-field scenario,
the amplitude structures of these channels are more similar.

\textbf{\textit{Case II}}:
The UE is located between the Rayleigh distances of the higher mmWave band and the lower sub-6GHz band.
Since the mmWave and Sub-6GHz channels experience the far-field and near-field scenarios, respectively,
their structural similarity is weaker compared to Case I.


It is worth mentioning that numerous studies have pointed out the similarities between the co-located mmWave and Sub-6GHz channel characteristics \cite{outofbandpdfPeter2016,Samimi2016TMTT,Kyosti2016EuCAP}.
In \cite{outofband2018}, the authors study the significant congruence between mmWave and Sub-6GHz channels,
suggesting that the out-of-band spatial information from one frequency band can be applied to assist
beam-selection and channel estimation in another frequency band in far-field scenarios.
Additionally, the out-of-band spatial information from the Sub-6GHz band is also leveraged to solve hybrid precoding \cite{LiAccess}, channel state information acquisition \cite{XiuAccess} and beam prediction \cite{MaKeVTC} problems in the mmWave band, respectively.
Consequently, the noteworthy similarities observed between the near-field angular-domain mmWave channel and the Sub-6GHz channel,
both in near-field and far-field scenarios, as illustrated in Fig.~\ref{Model},
can be effectively utilized to enhance the accuracy and reliability of mmWave channel estimation.
This aspect will be further investigated and studied in the subsequent analysis.

\subsection{Mathematical Formulation}
\label{S22}
At the $k$-th subcarrier and the $m$-th pilot transmission for one UE,
the relationship between the received signal $y_{m,k}$ and the near-field channel vector $\mathbf{h}_{k}$ can be expressed as
\begin{equation}
    y_{m,k}=\bm{\psi}_m\mathbf{h}_k+n_{m,k},
    \label{BaseModel}
\end{equation}
where $\mathbf{\psi}_m \in \mathbb{C}^{1 \times N}$ represents the pilots at the $m$-th pilot transmission,
and $n_{m,k} \in \mathbb{C}$ denotes the additive white Gaussian noise with zero mean and variance $\sigma^2$.
Based on (\ref{BaseModel}), the received signals of all $K$ subcarriers and $M$ pilot transmissions can form a joint matrix $\mathbf{Y} \in \mathbb{C}^{M \times K}$, and
\begin{equation}
    \mathbf{Y}=\mathbf{\Psi}\mathbf{H}+\mathbf{N},
    \label{Transmission}
\end{equation}
where $\mathbf{Y}_{m,k}=y_{m,k}$.
$\mathbf{\Psi}=[\mathbf{\psi}_1^{\rm H},\mathbf{\psi}_2^{\rm H},\cdots,\mathbf{\psi}_M^{\rm H}]^{\rm H} \in \mathbb{C}^{M \times N}$ is the joint pilot matrix and
$\mathbf{H}=[\mathbf{h}_1,\mathbf{h}_2,\cdots,\mathbf{h}_K] \in \mathbb{C}^{N \times K}$ is the channel matrix.
$\mathbf{N} \in \mathbb{C}^{M \times K}$ represents the additive noise matrix,
whose $(m,k)$-th entry is $n_{m,k}$, and is independently and identically distributed (i.i.d.) as $\mathcal{CN}(0,\sigma^2)$.

By exploiting the sparsity in the angular domain, $\mathbf{H}$ can be transformed as $\mathbf{H}=\mathbf{F}\mathbf{X}$ \cite{Block2},
where $\mathbf{F} \in \mathbb{C}^{N\times N}$ is the DFT codebook,
and the column vectors of $\mathbf{X}$ exhibit weak sparsity with block structure.
Therefore, by letting $\mathbf{\Psi}=\mathbf{A}\mathbf{F}^{\rm H}$, (\ref{Transmission}) can be simplified to
\begin{equation}
    \label{Problem}
    \mathbf{Y}=\mathbf{A}\mathbf{X}+\mathbf{N},
\end{equation}
where $\mathbf{A}\in \mathbb{C}^{M \times N}$ is called the measurement matrix,
and $\mathbf{A}$ can be a standard complex Gaussian matrix \cite{Scarletttsp2013}.
In the following, we elucidate the naturally occurring side information, including the structural characteristic information and the out-of-band spatial information,
used for estimating the angular-domain near-field channel $\mathbf{X} \in \mathbb{C}^{N \times K}$ in (\ref{Problem}), respectively.

\subsubsection{Structural Characteristic Information}
\label{S221}
Structural characteristic information indicates that the support set $\mathcal{S}$ for different subcarriers is typically the same for different block-sparse column vectors in $\mathbf{H}$ \cite{Support}.
We denote $S$ as the size of $\mathcal{S}$, and the sparsity is defined as $\frac{S}{N}$.
Therefore, the angular-domain channel $\mathbf{X}$ has similar structure as $\mathbf{H}$,
which is referred to as the Block Multiple-Measurement Vector (BMMV) feature \cite{MMV}
and is also depicted on the right-hand side in Fig.~\ref{Sparsity}.
As a result, $\mathbf{X}$ can be written as $[\mathbf{X}_1^{\rm H},\mathbf{X}_2^{\rm H},\cdots,\mathbf{X}_{N/d}^{\rm H}]^{\rm H}$,
and $\mathbf{A}=[\mathbf{A}_1,\mathbf{A}_2,\cdots,\mathbf{A}_{N/d}]$,
where $\mathbf{X}_i$ and $\mathbf{A}_j$ represents the $i$-th row-block submatrix in $\mathbf{X}$ and the $j$-th column-block submatrix in $\mathbf{A}$, respectively,
and $d$ denotes the block length of the BMMV feature.
When $d=1$, submatrices $\mathbf{X}_i$ and $\mathbf{A}_j$ degenerate into vectors $\mathbf{x}_i$ and $\mathbf{a}_j$ naturally.

\subsubsection{Out-of-band Spatial Information}
\label{S222}
In addition, the authors in \cite{outofband2018} demonstrate that the out-of-band spatial information extracted from the Sub-6GHz band
can be leveraged in the mmWave channel estimation.
Based on the channel reciprocity in TDD mode and the channel estimation method in \cite{Minn2006TCOM,outofband2018},
the channel matrix $\underline{\mathbf{H}}\in \mathbb{C}^{\underline{N} \times K}$ in the Sub-6GHz band can be obtained by uplink channel estimation.
Moreover, by leveraging the Sub-6GHz codebook $\underline{\mathbf{F}} \in \mathbb{C}^{\underline{N} \times N}$,
which consists of the Sub-6GHz antenna array response vectors sampled at the same spatial points used for mmWave codebook $\mathbf{F}$,
the joint angular-domain channel matrix $\underline{\mathbf{X}}\in \mathbb{C}^{N \times K}$ from the Sub-6GHz system with lower-frequency carrier is available.
To be specific, $\mathbf{\underline{F}}$ is set to $[\mathbf{\underline{F}}]_{mn}=e^{j2\pi\frac{(m-1)(n-1)}{N}}/\sqrt{N}$ according to \cite{dai2022tcom},
and $\mathbf{\underline{X}}$ can therefore be obtained by $\mathbf{\underline{X}}=\mathbf{\underline{F}}^H\mathbf{\underline{H}}$,
satisfying $\underline{\mathbf{H}}=\underline{\mathbf{F}}\underline{\mathbf{X}}$.
Consequently, we can generate the out-of-band probability vector $\mathbf{p}$ to assist the estimation of $\mathbf{X}$ in the mmWave band,
where $p_i$, the $i$-th entry of $\mathbf{p}$, denotes the nonzero probability of $\mathbf{X}_i$ to be nonzero.

In general, due to lack of scattering and diffraction,
the Sub-6GHz angular-domain channel $\underline{\mathbf{X}}$ usually does not exhibit sparsity structure like $\mathbf{X}$.
However, as observed in Fig.~\ref{Model}, the central points of the significant paths in both mmWave and Sub-6GHz bands have similar positions.
As a result, similarities exist in the amplitude between the entries of submatrices in the mmWave $\mathbf{X}$ and Sub-6GHz $\underline{\mathbf{X}}$ \cite{outofband2018},
where the entries of $\underline{\mathbf{X}}$ corresponding to the nonzero entries in $\mathbf{X}$
tend to have a larger amplitude.
Therefore, based on the structural feature of the mmWave channel $\mathbf{X}$,
the Sub-6GHz channel $\underline{\mathbf{X}}$ could also be segmented with the same block length $d$ correspondingly,
i.e., $\underline{\mathbf{X}}=[\underline{\mathbf{X}}_1^{\rm{H}},\underline{\mathbf{X}}_2^{\rm{H}},\cdots,\underline{\mathbf{X}}_{N/d}^{\rm{H}}]$,
providing effective out-of-band information for the assistance of mmWave channel estimation.
Formulatically, $p_i$ of the out-of-band probability vector $\mathbf{p} \in [0,1]^{N/d}$ can be calculated as
\begin{equation}
    p_i=f(\|\underline{\mathbf{X}}_i\|_F)=f(\underline{x}_i),
    \label{PriorVec}
\end{equation}
where $\|\underline{\mathbf{X}}_i\|_F$ is denoted as $\underline{x}_i$,
and $f$ is a determined function mapping the Frobenius norm of $\underline{\mathbf{X}}_i$ to a scalar within the range of (0,1) as the probability.
It is worth mentioning that, $f$ should be selected as a monotonic increasing function \cite{Scarletttsp2013}.

Note that the out-of-band probability $\mathbf{p}$ depends on the amplitude similarity between $\mathbf{X}$ and $\underline{\mathbf{X}}$.
Therefore, the Sub-6GHz angular-domain matrix $\underline{\mathbf{X}}$,
which represents a coarse estimation of the channel angular information,
can be applied to provide out-of-band spatial information through $\mathbf{p}$
for the recovery of the mmWave angular-domain matrix $\mathbf{X}$
even if the structural features of $\mathbf{X}$ and $\underline{\mathbf{X}}$ are not completely consistent.

Consequently, by taking the Frobenius norm of the residual error matrix $\mathbf{Y}-\mathbf{A}\mathbf{X}$ as the optimization object function,
the side information-assisted channel estimation problem in the considered system can be written as
\begin{equation}
    \begin{aligned}
        \label{Optim}
         & \min\limits_{\mathbf{X}} \|\mathbf{Y}-\mathbf{A}\mathbf{X}\|_F \\
         & s.t. \quad supp(\mathbf{X})=S,
    \end{aligned}
\end{equation}
given that $\mathbf{X}$ exhibits block sparsity with length $d$, and the out-of-band spatial information can be expressed as
\begin{equation}
    P(\mathbf{X}_i \neq \mathbf{0})=p_i.
\end{equation}

\emph{Remark 1}:
For Sub-THz near-field communication systems with higher carrier frequency,
it can be observed in Fig.~\ref{AoD_THz} that,
the channel characteristics, like the AoDs of different frequency bands,
remain congruent although the carrier frequency increases drastically.
Therefore, the aforementioned amplitude similarities between still exist between different Sub-THz bands,
and can be utilized to assist reliable channel estimation in the higher Sub-THz frequency band,
suggesting that our proposed system model and methodologies are also applicable to the imperative Sub-THz scenarios
in near-field communications.

    {
        \begin{figure}[tp!]
            \begin{center}
                \includegraphics[width=0.5\textwidth]{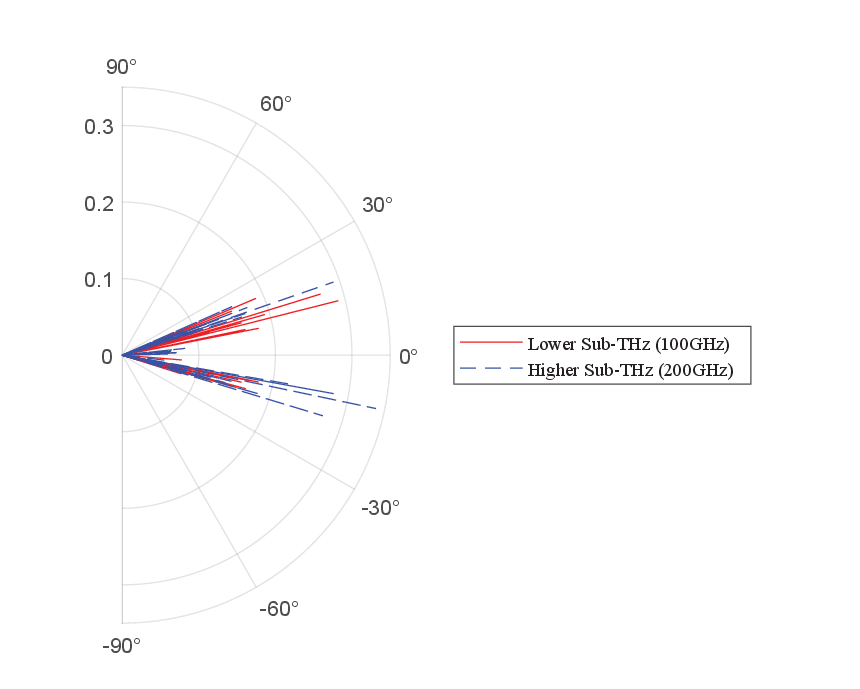}
            \end{center}
            \caption{The AoDs in the dual-band Sub-THz near-field communication system.}
            \label{AoD_THz}
        \end{figure}
    }

\section{Dual-Band Channel Estimation} \label{S3}
In this section, we start with the CLW-OMP algorithm and then propose the CSLW-BOMP algorithm for channel estimation in the elaborating dual-band XL-MIMO systems.
In these algorithms, a prior factor $v(p_i)$ is derived to guarantee accurate selection of nonzero channel-taps
based on the out-of-band probability vector $\mathbf{p}$ stemming from the lower-frequency communication system.
The single subcarrier and multiple subcarrier cases are considered seperately,
since the degree of freedom of the involved non-central $\chi^2$ distributions increases with the number of subcarriers,
and the characteristic function method applied in single subcarrier case is not suitable for more complicated multiple carrier cases with higher degrees of freedom.
Moreover, for the off-grid scenarios, CSLW-OMP is proposed as an effective solution,
which is analyzed as the corollary of the CSLW-BOMP algorithm.

\subsection{Single Subcarrier Case}
\label{S31}
Consider the case where $K=d=1$,
and the problem (\ref{Problem}) can thus be simplified into
\begin{equation}
    \mathbf{y}=\mathbf{A}\mathbf{x}+\mathbf{n},
\end{equation}
where $\mathbf{y}$, $\mathbf{x}$ and $\mathbf{n}$ are the vector form of
the joint receiving matrix $\mathbf{Y}$, the angular-domain channel matrix $\mathbf{X}$ and the additive noise matrix $\mathbf{N}$, respectively.

In the conventional OMP approach, the best support index is selected by the maximum $|\mathbf{a}_i^{\rm H}\mathbf{y}|$ with respect to $i$ in each iteration step,
which is equivalent to solving $\max\limits_{i} {|\mathbf{a}_i^{\rm H}\mathbf{y}|^2}$.
However, when the out-of-band probability vector $\mathbf{p}$ is achievable,
the CLW-OMP algorithm can be proposed to further exploit this out-of-band spatial information,
in which $|\mathbf{a}_i^{\rm H}\mathbf{y}|^2+v(p_i)$ is maximized instead of $|\mathbf{a}_i^{\rm H}\mathbf{y}|$ to select the optimal support index in each iteration step.
Here, $v(p_i)$ is a prior factor based on $\mathbf{p}$ and other scalar parameters,
serving as a penalty to introduce out-of-band spatial information to the conventional OMP algorithm.
For the optimal $v(p_i)$, we provide the following theorem.

\begin{theorem}
    \label{Thm31}
    (CLW-OMP): Among all choices of $v(p_i)$, the following expression
    \begin{equation}
        v(p_i)=D\ln\Bigl(\frac{p_i}{1-p_i}\Bigr), i=1,2,\cdots,N,
    \end{equation}
    is optimal in minimizing the error probability of wrongly choosing a nonzero channel-tap over a zero one in $\mathbf{x}$ in each iteration of CLW-OMP.
    The coefficient $D$ can be calculated as
    \begin{equation}
        \begin{aligned}
             & \sigma_1^2=\frac{M}{2}((S-1)g^2+\sigma^2),                                                                                       \\
             & \sigma_2^2=\frac{M}{2}(Sg^2+\sigma^2),                                                                                           \\
             & A=\frac{1}{\sigma_2^2},                                                                                                          \\
             & B=\frac{1}{2}e^{-\frac{M^2g^2\sigma_2^2}{2\sigma_1^2(\sigma_1^2+\sigma_2^2)}}\frac{\sigma_1^2+\sigma_2^2}{\sigma_1^2\sigma_2^2}, \\
             & D=\frac{1}{A-B}.
        \end{aligned}
        \label{D1}
    \end{equation}
\end{theorem}

\begin{IEEEproof}
    See Appendix A.
\end{IEEEproof}
In practical scenarios, $g^2$ is usually far less than $M$ and $S$,
leading to $\sigma_1^2 \approx \sigma_2^2$, which is denoted as $\sigma_0^2$.
Therefore, the coefficient $D$ will take a much simpler form when applied.
\begin{equation}
    D=\frac{\sigma_0^2}{1-e^{-\frac{M^2g^2}{4\sigma_0^2}}} \approx \frac{4\sigma_0^4}{M^2g^2}.
    \label{Dsim1}
\end{equation}

\emph{Remark 2}:
Compared to \cite{Scarletttsp2013}, the coefficient $D$ in (\ref{D1}) is more generalized,
since complex signals in practical communication scenarios are considered.
In addition, $D$ derived from Theorem \ref{Thm31} is more accurate than that in \cite{outofband2018},
where the noise variance $\sigma^2$ is taken as the coefficient in the prior factor $v(p_i)$ without theoretical analysis in the formula of \cite{outofband2018}.

\subsection{Multiple Subcarrier Case}
\label{S32}
In multiple subcarrier cases where $K>1$, the system model now follows the form of (\ref{Problem}),
and $\mathbf{X}$ satisfies the BMMV feature where $d>1$.
Correspondingly, the correlation term in Section~\ref{S31} changes from the modulus $|\mathbf{a}_i\mathbf{y}|^2$ to the Frobenius norm $\|\mathbf{A}_i\mathbf{Y}\|_F^2$.

Under this circumstance, the correlation term, which follows the non-central $\chi^2$ distribution, will have a higher degree of freedom,
leading to an additional complicated factor in the characteristic function, which can not be solved easily by the method leveraged in Theorem \ref{Thm31}.
Consequently, the Patnaik's second moment approximation, which is suitable for estimating non-central $\chi^2$ distributions with high degrees of freedom, is applied \cite{Patnaik},
and the optimal analytical expression of the prior factor $v(p_i)$ can therefore be acquired,
which is presented in the following theorem.

\begin{theorem}
    \label{Thm32}
    (CSLW-BOMP):
    Among all choices of $v(p_i)$,
    \begin{equation}
        v(p_i)=D\ln\Bigl(\frac{p_i}{1-p_i}\Bigr), i=1,2,\cdots,N,
    \end{equation}
    minimizes the error probability of wrongly selecting a nonzero channel-tap over a zero one in $\mathbf{x}$ in each iteration of CSLW-BOMP,
    where the coefficient $D$ follows
    \begin{equation}
        \begin{aligned}
            \sigma_1^2 & =M(Sdg^2+\sigma^2),                                                         \\
            \sigma_2^2 & =M[(Sd-1)g^2+\sigma^2],                                                     \\
            \rho       & =\frac{2dK+\frac{4dKM^2g^2}{\sigma_2^2}}{2dK+\frac{2dKM^2g^2}{\sigma_2^2}}, \\
            \beta_1    & =\frac{1}{\rho\sigma_2^2},                                                  \\
            \beta_2    & =\frac{1}{\sigma_1^2},                                                      \\
            D          & =\frac{1}{\beta_2-\beta_1}.
        \end{aligned}
    \end{equation}
    \label{D2}
\end{theorem}

\begin{IEEEproof}
    See Appendix B.
\end{IEEEproof}

Similarly, $D$ can be simplified into
\begin{equation}
    D=\frac{\rho}{\rho-1}\sigma_0^2=\frac{\sigma_0^2+2M^2g^2}{M^2g^2}\sigma_0^2,
    \label{Dsim2}
\end{equation}
where $\sigma_0^2=M(Sdg^2+\sigma^2)$.
Moreover, when $d>1$, the relative difference between $\sigma_1^2$ and $\sigma_2^2$ with respect to their specific value becomes smaller with the increase of $d$,
which results in smaller error in the approximation $\sigma_2^2\approx \sigma_1^2=\sigma_0^2$ and leads to more accurate estimation in (\ref{Dsim2}).
The CSLW-BOMP algorithm is summarized in Algorithm \ref{LWBOMP}.

Furthermore, in (\ref{Dsim2}), the coefficient $D$ is inversely proportional to the signal-to-noise ratio (SNR).
To be specific, when the SNR is low, the coefficient $D$ becomes large accordingly,
which leads to a larger proportion of the prior factor $v(p_i)$ in the index selection mechanism per iteration.
In contrast, when the system SNR is high, $D$ tends to approach a small value and $v(p_i)$ is relatively small compared to the correlation term $\|\mathbf{A}_i\mathbf{Y}\|_F^2$,
resulting in the CSLW-BOMP algorithm to converge to the conventional BOMP algorithm.

When $d=1$, the CSLW-BOMP algorithm becomes the CSLW-OMP algorithm,
which is presented in the following corollary.
The CSLW-OMP algorithm can be applied in off-grid scenarios,
where the CSLW-BOMP algorithm has difficulties in handling the off-grid structure.

\begin{corollary}
    \label{Thm33}
    (CSLW-OMP): Among all $v(p_i)$, the optimal choice minimizing the error probability can be expressed as
    \begin{equation}
        v(p_i)=D\ln\Bigl(\frac{p_i}{1-p_i}\Bigr), \quad i=1,2,\cdots,N,
    \end{equation}
    where the coefficient $D$ can be calculated by setting $d=1$ in (\ref{D2}).
\end{corollary}

The coefficient $D$ can also be simplified by utilizing the fact that $\sigma_1^2 \approx \sigma_2^2 =\sigma_0^2$ in practical scenarios, i.e.,
\begin{equation}
    D=\frac{\sigma_0^2+2M^2g^2}{M^2g^2}\sigma_0^2.
    \label{Dsim3}
\end{equation}
It is clear that (\ref{Dsim3}) is a special case for (\ref{Dsim2}) when $d=1$,
and the algorithm design of CSLW-OMP can be acquired by letting $d=1$ in Algorithm~\ref{LWBOMP},
which is omitted here.

\emph{Remark 3}:
The CLW-BOMP algorithm can also be analyzed based on Theorem \ref{Thm32} by setting $K=1$ instead of $d$,
which is similar to the CSLW-OMP algorithm.
Since the block structure in the single subcarrier case also involves non-central $\chi^2$ distributions with high degrees of freedom,
the Patnaik's second moment approximation can be utilized to obtain the optimal $v(p_i)$ in the CLW-BOMP algorithm
similar to the proof of Theorem \ref{Thm32}.
{
\begin{algorithm}[tp!]
    \caption{CSLW-BOMP Algorithm}
    \begin{algorithmic}[1]
        \REQUIRE $\mathbf{Y}$, $\mathbf{A}$, $g$, $\sigma^2$, block length $d$, out-of-band probability vector $\mathbf{p}$, convergence limit $S$
        \ENSURE Channel estimation $\hat{\mathbf{X}}$
        \STATE \textbf{Initialize:} $i=1$, $\mathbf{R}=\mathbf{0}$, $\mathcal{S}=\emptyset$
        \STATE \textbf{repeat}
        \STATE Calculate $D$ with $K$ in (\ref{D2}) replaced by $K-i+1$
        \STATE Solve $\arg\max\limits_{k \notin \mathcal{S}}\|\mathbf{A}_i^{\rm H}\mathbf{Y}\|_F^2+D\ln{\frac{p_k}{1-p_k}}$ to obtain the optimal block index $k_i$ for thr $i$-th iteration
        \STATE Update the support set $\mathcal{S}=\mathcal{S} \bigcup \{(k_i-1)d+1, (k_i-1)d+2, \cdots, k_id\}$
        \STATE Update the channel estimation $\hat{\mathbf{X}}=\arg\min\limits_{\hat{\mathbf{X}}}\|\mathbf{Y}-A_{\mathcal{S}}\hat{\mathbf{X}}\|_F$ to obtain new $\hat{\mathbf{X}}$
        \STATE Update the residual matrix $\mathbf{R}=\mathbf{Y}-\mathbf{A}_{\mathcal{S}}\hat{\mathbf{X}}$
        \STATE $i=i+1$
        \STATE \textbf{until} $|\mathcal{S}|=S$
    \end{algorithmic}
    \label{LWBOMP}
\end{algorithm}
}

\emph{Remark 4}:
Theorem \ref{Thm32} assumes that $\mathbf{X}$ has a on-grid block sparse structure,
where the nonzero blocks consistently appear in units of $d$.
In off-grid scenarios, the CSLW-OMP proposed in Corollary~\ref{Thm33},
which can be realized by setting $d=1$ in the CSLW-BOMP described in Algorithm \ref{LWBOMP}, can be well applied,
since it takes the off-grid structure where $d>1$, as on-grid structure with $d=1$.
Consequently, the CSLW-OMP algorithm is capable of selecting the support indices one by one,
therefore avoiding the support misalignment casued by leveraging the block structure.
Additionally, when applying the proposed CSLW-BOMP algorithm proposed in Theorem \ref{Thm32} in off-grid scenarios,
it is possible to gradually decrease the value of $d$ in order to better adapt to the off-grid structure,
leading to improved recovery performance and higher computational complexity.
If $d$ eventually decreases to 1, the CSLW-BOMP algorithm will converge to the CSLW-OMP algorithm,
which is specifically designed to handle this off-grid structure.

\section{Simulation Results} \label{S4}
In our simulation, the dual-band XL-MIMO communication system with $K=32$ subcarriers is considered.
Following the antenna settings in \cite{Yuan2023TAP,Wang2024TWC,Tang2024TCOM,Yang2023CL,Nayir2022VTC},
the number of antennas at the BS side is set to $N=256$ on the $f=28$GHz mmWave band,
and $\underline{N}=32$ on the $\underline{f}=3.5$GHz Sub-6GHz band.
Quantitatively, the antenna aperture can then be calculated as $\frac{Nc}{2f}=1.37$m,
which is acceptable in practical use.
For the generation of the out-of-band channel matrix $\underline{\mathbf{H}}$,
each entry in $\underline{\mathbf{H}}$ can be obtained by multiplying the corresponding entry in $\mathbf{H}$ by a specific coefficient $Q$,
whose amplitude and phase depend on the relationship of the central frequencies between the mmWave band and the Sub-6GHz band,
which is denoted as $f_m$ and $f_s$, respectively \cite{outofband2018}.

Specifically, with the assumption that the support set $\mathcal{S}$ is the same for $\mathbf{X}$ and $\underline{\mathbf{X}}$,
the coefficient $Q$ for nonzero blocks in $\underline{\mathbf{X}}$ can be calculated as
\begin{equation}
    \begin{aligned}
        \gamma  & =\frac{|f_m-f_s|}{\max(f_m,f_s)}, \\
        |Q|     & = \gamma R_1 \delta,              \\
        \arg(Q) & =2\pi \gamma R_2 \delta,          \\
    \end{aligned}
\end{equation}
where $R_1$ and $R_2$ are two independent variables, whose values are $\pm 1$ with equal probability,
and $\delta$ is uniformly distributed between 0 and 1.
Moreover, the channel-taps in $\underline{\mathbf{X}}$ in zero row-block submatrices are set to follow the distribution of $\mathcal{CN}(0,\sigma_{n}^2)$ to simulate the perturbations on the zero channel-taps in $\underline{\mathbf{X}}$
due to the difference in frequencies between $f_s$ and $f_m$,
and $\sigma_n^2=\frac{\gamma^2}{C}$ \cite{outofband2018}.
$C$ is the variance amplitude ratio for nonzero channel-taps and zero channel-taps in $\underline{\mathbf{X}}$.

Once the Sub-6GHz support channel $\underline{\mathbf{X}}$ is generated, the out-of-band probability vector $\mathbf{p}$ can then be acquired based on (\ref{PriorVec}).
The function $f$ is selected as
\begin{equation}
    p_i=f({\underline{x}_i})=\frac{{\underline{x}_i}-\underline{x}_{min}}{\underline{x}_{max}-\underline{x}_{min}},
    \label{f}
\end{equation}
where $\underline{x}_{max}$ and $\underline{x}_{min}$ represent the maximum and minimum in $\{\underline{x}_1,\underline{x}_2,\cdots,\underline{x}_{N/d}\}$, respectively,
and (\ref{f}) can be leveraged to generate $\mathbf{p}$ in both Case I and Case II as illustrated in Section~\ref{S21}.
Meanwhile, we consider the more general Case I scenario,
where both mmWave and Sub-6GHz bands are within the near-field region,
leading to stronger block structure for both channels.
The far-field channel, i.e., the Sub-6GHz system in Case II,
is regarded as a special case by setting the block length as 1 or a smaller value
compared with the near-field scenario.
According to (\ref{f}), $p_i$ of the row-block submatrix with the maximum Frobenius norm in $\underline{\mathbf{X}}$ is normalized to 1 in $\mathbf{p}$,
and therefore has the value of infinity under the influence of the prior factor $v(p_i)=D\log(\frac{p_i}{1-p_i})$.
As a result, the CSLW-BOMP algorithm always chooses the row-block submatrix in $\mathbf{X}$ of the mmWave band
which corresponds to the strongest row-block submatrix in the support channel $\underline{\mathbf{X}}$ of the Sub-6GHz system,
as the starting support information in the first iteration for recovering $\mathbf{X}$,
leading to potential performance enhancement of the algorithm.
Similarly, the row-block submatrix with the minimum Frobenius norm in $\underline{\mathbf{X}}$ becomes negative infinity due to the existence of $v(p_i)$,
which improves the overall performance of our proposed algorithm by consistently ignoring this worst-performing submatrix.
The detailed simulation parameters are listed in Table~\ref{Param}.

\begin{figure}[tp!]
    \begin{center}
        \includegraphics[width=0.5\textwidth]{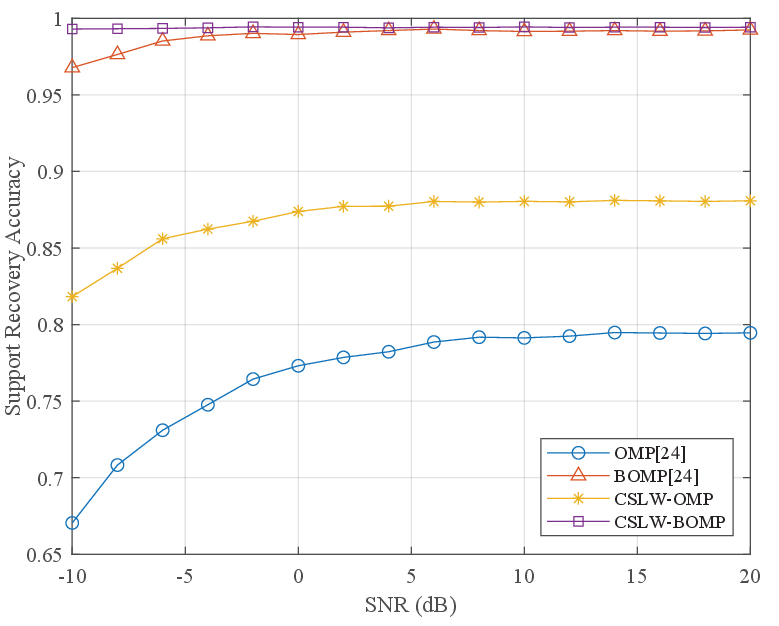}
    \end{center}
    \caption{Support Recovery Accuracy versus SNR.}
    \label{Acc}
\end{figure}

First, numerical results are provided in Fig.~\ref{Acc} to backup analytical results in Theorems \ref{Thm31}, \ref{Thm32} and Corollary \ref{Thm33}.
Since better prior factor $v(p_i)$ leads to more assured results in minimizing the error probability of wrongly choosing a nonzero channel-tap over a zero one in $\mathbf{X}$,
as described in Theorems \ref{Thm31}, \ref{Thm32} and Corollary \ref{Thm33},
the support recovery accuracy is utilized as the measuring metric to showcase the channel-tap selection performance.
Let $M=25$ and $\frac{S}{d}=5$.
As shown in Fig.~\ref{Acc}, our proposed CSLW-BOMP algorithm outperforms its counterparts,
while the performance of the CSLW-OMP algorithm is also better than that of the conventional OMP algorithm,
which proves the effectiveness of our derived prior factor $v(p_i)$ in leveraging the out-of-band spatial information.
In addition, with the help of the out-of-band spatial information utilized by $v(p_i)$,
the CSLW-BOMP algorithm is able to estimate the support of the mmWave channel matrix $\mathbf{X}$ more accurately when the SNR is low.
At high SNRs, the support recovery accuracy of CSLW-BOMP and conventional BOMP algorithms tends to be the same,
since the coefficient $D$ reaches a small value and the CSLW-BOMP slowly converges to the BOMP algorithm as discussed in Theorem \ref{Thm32}.

As for the benchmarks in the performance comparison,
we adopt various channel estimation methods based on greedy algorithms,
beam training and low-rank matrix completion (LRMC)  theory, which are listed below.
1) \emph{OMP} and \emph{BOMP}: Conventional greedy CS algorithms without any side information \cite{Block1}.
2) \emph{Weighted beam recovery (WBR)}: The beam selection-based method in \cite{outofband2018} using only out-of-band spatial information.
3) \emph{Probability-based beam selection (PBS)}: The beam training based channel estimation algorithm in \cite{Choi2020TCOM}.
4) \emph{Fast alternative least squares (FALS)}: The two-stage channel estimation algorithm based on low-rank matrix completion theory in \cite{Kim2022ICASSP}.
5) \emph{Least squares (LS)}: Conventional least squares channel estimation, where $\hat{\mathbf{X}}=\mathbf{\Psi}^{\dagger}\mathbf{Y}$.

\begin{table}[tp!]
    \begin{center}
        \caption{Parameter Settings.}
        \label{Param}
        \setlength{\tabcolsep}{2mm}
        \begin{tabular}{cc}
            \toprule[0.8pt]
            Parameters                        & Values \\
            \toprule[0.8pt]
            Support bandwidth $BW$            & 150MHz \\
            Object bandwidth $\underline{BW}$ & 850MHz \\
            Block length $d$                  & 4      \\
            Amplitude ratio $C$               & 3      \\
            Number of iteration $N_{iter}$    & 1000   \\
            \toprule[0.8pt]
        \end{tabular}
    \end{center}
\end{table}

\begin{figure}[tp!]
    \begin{center}
        \includegraphics[width=0.5\textwidth]{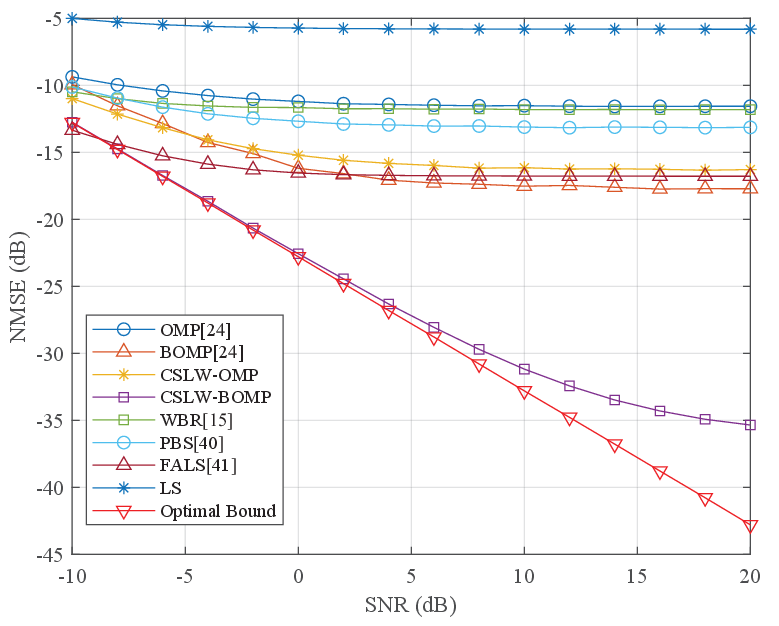}
    \end{center}
    \caption{On-grid NMSE performance versus SNR.}
    \label{NMSE}
\end{figure}

In Fig.~\ref{NMSE}, the channel estimation performance in on-grid scenarios is compared,
with the number of row-block matrices $\frac{S}{d}$, number of pilot transmission $M$ fixed to 5 and 25 respectively.
The normalized mean-square error (NMSE) is leveraged as the criterion, which can be expressed as
\begin{equation}
    \text{NMSE}=\frac{\|\hat{\mathbf{X}}-\mathbf{X}\|_F^2}{NK},
\end{equation}
where $\mathbf{X}$ and $\hat{\mathbf{X}}$ represent the angular-domain channel matrix and the estimated angular-domain channel, respectively.
Moreover, the optimal bound where the channel-tap selections are set to be correct in each iteration is adopted as the upper bound in on-grid scenarios.
It can be observed that our proposed CSLW-BOMP algorithm outperforms all the other benchmarks,
whose NMSE performance is close to that of the optimal bound benchmark,
while the CSLW-OMP algorithm also has an advantage in comparison to its conventional counterparts,
indicating the effectiveness of the application of out-of-band spatial information.
Moreover, the NMSE performance of BOMP and CSLW-BOMP is also superior to that of the OMP and CSLW-OMP algorithms,
since the BOMP and the CSLW-BOMP algorithms take the structural characteristic information into consideration and fully utilize the block structure in $\mathbf{X}$,
which efficiently enhances the robustness of the algorithm, preventing severe nonzero channel-tap selection errors for assured estimation performance.
In addition, the exploitation of both types of side information leads to the continuous decrease of the NMSE of the CSLW-BOMP algorithm as the SNR increases,
while other algorithms compared tend to reach a plateau at high SNR scenarios.

\begin{figure}[tp!]
    \begin{center}
        \includegraphics[width=0.5\textwidth]{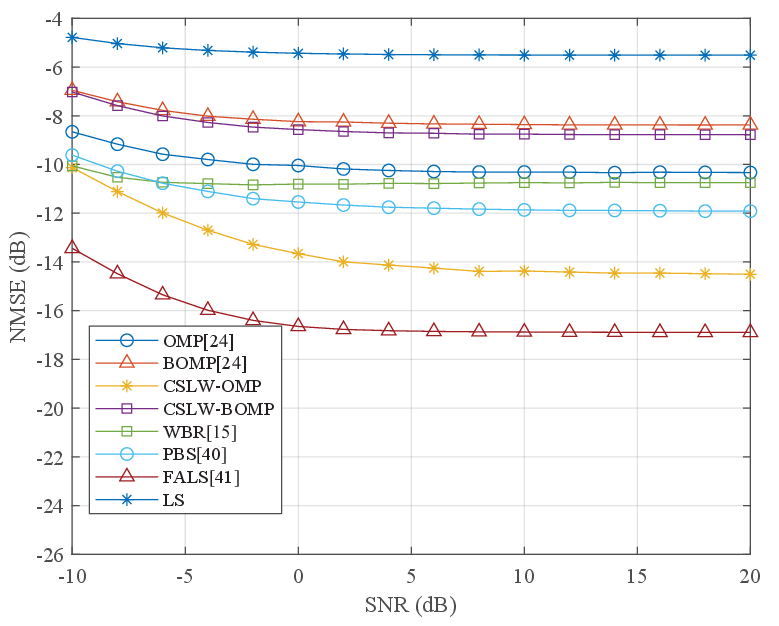}
    \end{center}
    \caption{Off-grid NMSE performance versus SNR.}
    \label{NMSE2}
\end{figure}

As illustrated in Fig.~\ref{NMSE2}, when the channel matrix $\mathbf{H}$ changes from on-grid to off-grid,
the NMSE performance of various algorithms varies drastically.
Although the channel grid cannot align with the block structure in off-grid scenarios,
leading to inherent defect in handling off-grid block structure,
the CSLW-BOMP still performs the best between the two block structure-based algorithms due to the assistance of out-of-band spatial information.
Meanwhile, the CSLW-OMP algorithm outperforms most of the benchmarks except the FALS method.
This is because the OMP algorithms can effectively adapt to the off-grid channel features
by treating the off-grid structure, where $d>1$ as on-grid features with $d=1$,
as elucidated in Remark 4.
Therefore, the CSLW-OMP algorithm can facilitate accurate estimation of the mmWave channel
by exploiting the sophisticated out-of-band spatial information from the Sub-6GHz band.
Regarding the two-stage FALS benchmark, in order to recover the whole $\mathbf{X}$,
certain entries in the channel matrix $\mathbf{X}$ need to be measured as prior knowledge.
Since $\mathbf{X}$ is typically not a low-rank matrix, the amount of the required pre-measurements is enormous,
leading to unacceptable measurement cost.
Numerically, in our settings, at least 65\% of the entries in $\mathbf{X}$ need to be measured in the first stage to perform reliable LRMC in the second stage,
making the FALS method impractical although it achieves a better performance of about 3dB compared to our proposed CSLW-OMP algorithm.

\begin{figure}[tp!]
    \begin{center}
        \includegraphics[width=0.5\textwidth]{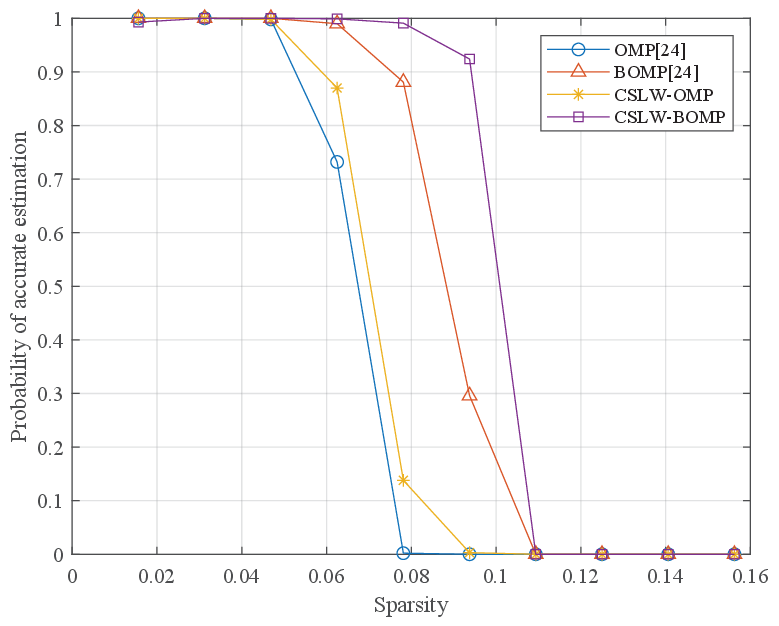}
    \end{center}
    \caption{Probability of accurate estimation versus sparsity.}
    \label{Sparsitylevel}
\end{figure}

In Fig.~\ref{Sparsitylevel}, the on-grid recovery performance of all the algorithms is presented
as the number of nonzero blocks $\frac{S}{d}$ increases, which is in proportion to the sparsity $\frac{S}{N}$.
We set $M=50$ and $\text{SNR}=10\text{dB}$.
The probability of accurate estimation, which is defined as the frequencies of the NMSE to be less than a threshold $\theta$,
is utilized to provide their performance, and $\theta=10^{-2}$.
It can be concluded that all the algorithms achieve accurate channel estimation with probability of accurate estimation as $100\%$ when the sparsity is low.
When the sparsity approaches 0.13, the probability of accurate estimation decreases,
but the CSLW-BOMP and CSLW-OMP algorithms still outperform other algorithms without out-of-band spatial information.
Furthermore, as the sparsity gradually increases, CSLW-BOMP leveraging both types of side information, i.e., structural characteristic information and out-of-band spatial information,
enjoys the best recovery performance,
being able to accurately recover the channel matrix even if the sparsity is larger than 0.1.
Hence, our proposed CSLW-BOMP algorithm can effectively improve the upper bound of the reconstructible sparsity,
and therefore acts as a stunning solution to the weak sparsity problem in near-field XL-MIMO systems.
Moreover, the proposed CSLW-OMP provides more assured results than those of the OMP algorithm,
which even approaches or exceeds the estimation performance of the BOMP assisted by the block structure,
and the OMP has the worst estimation performance because of the most restricted upper bound of reconstructible sparsity.

\begin{figure}[tp!]
    \begin{center}
        \includegraphics[width=0.5\textwidth]{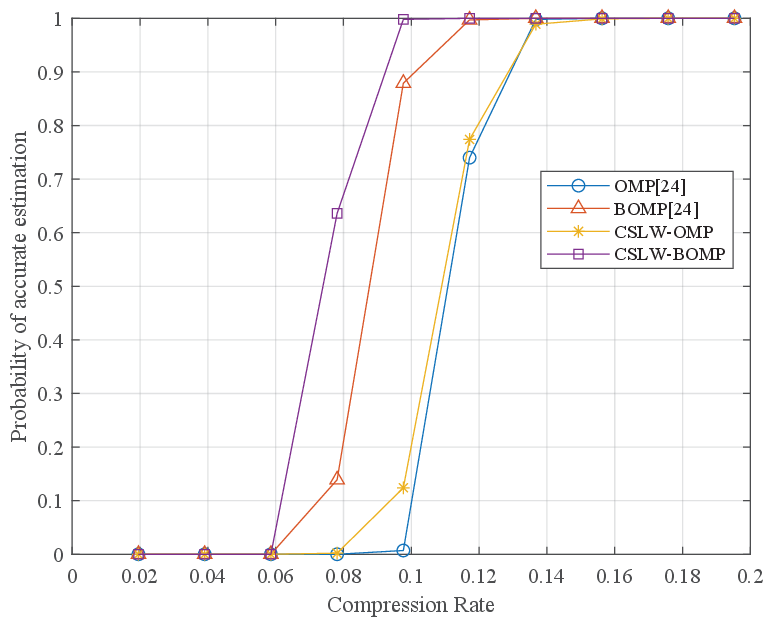}
    \end{center}
    \caption{Probability of accurate estimation versus compression rate.}
    \label{Compression}
\end{figure}

In Fig.~\ref{Compression}, the probability of accurate estimation is simulated as a function of the compression rate $\frac{M}{N}$,
which is proportional to the number of required pilot transmissions $M$.
It unveils that the probability of accurate estimation increases with the increase of compression rate,
and the CSLW-BOMP still enjoys the highest probability of accurate estimation under the same compression rate among all compared algorithms,
which is similar to the conclusions in Fig.~\ref{Sparsitylevel}.
Meanwhile, when the probability of accurate estimation is fixed,
CSLW-BOMP requires the least pilot transmissions, followed by BOMP, CSLW-OMP and OMP.
For $100\%$ probability of accurate estimation,
CSLW-BOMP can reduce the pilot overhead by 25\% when compared to the BOMP algorithm,
and $30.1\%$ when compared to the OMP algorithm,
substantiating the feasibility of the proposed CSLW-BOMP in low-overhead dual-band XL-MIMO communications.

\begin{figure}[tp!]
    \begin{center}
        \includegraphics[width=0.5\textwidth]{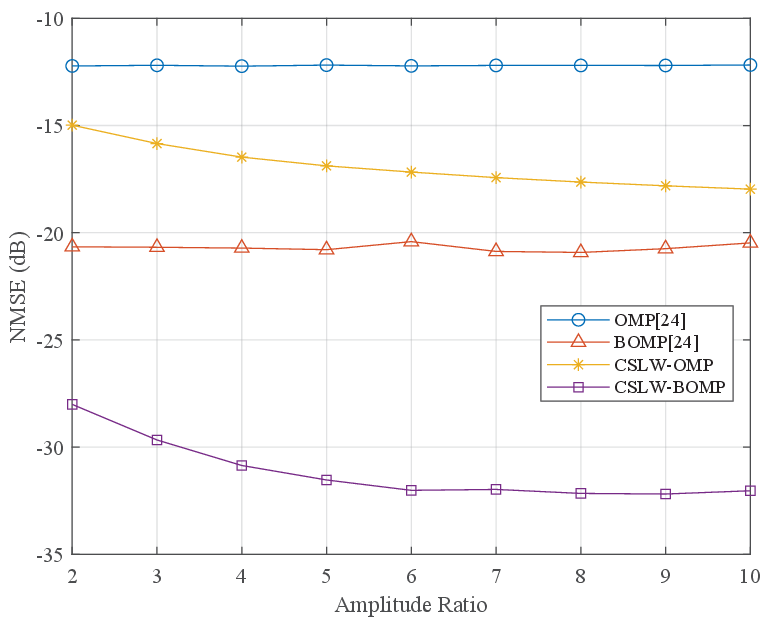}
    \end{center}
    \caption{NMSE versus Amplitude Ratio $C$.}
    \label{Perturbation}
\end{figure}

Finally, the influence of the perturbation level of the Sub-6GHz channel,
which is represented by the amplitude ratio $C$,
on the channel estimation performance is simulated in Fig.~\ref{Perturbation},
with $M$, $\frac{S}{d}$ and $\text{SNR}$ fixed to $25$, $5$ and $10\text{dB}$, respectively.
From Fig.~\ref{Perturbation},
we can know that the estimation performance of the proposed CSLW-BOMP and CSLW-OMP
improves with the increase of the amplitude ratio $C$,
as the out-of-band spatial information extracted from the support Sub-6GHz band is becoming more accurate.
Since the out-of-band spatial information from $\mathbf{\underline{X}}$ is not leveraged in the conventional OMP and BOMP benchmarks,
the performance of the OMP and BOMP algorithms is not affected by the amplitude ratio $C$.
Thus, the NMSE of the OMP and BOMP algorithms remains unchanged as $C$ increases.
Additionally, although the out-of-band spatial information could be incorrect due to perturbation,
the proposed CSLW-BOMP algorithm still exhibits superior performance when compared to other benchmarks,
while the CSLW-OMP algorithm also outperforms the conventional OMP algorithm.
Numerically, when $C=2$, suggesting that the variance of channel-taps in $\underline{\mathbf{X}}$ corresponding to the zero channel-taps in $\mathbf{X}$
is half that of nonzero channel-taps,
the CSLW-BOMP algorithm has a gain of about 7dB in NMSE when compared to the BOMP benchmark.
Meanwhile, the CSLW-OMP algorithm has a gain of about 3dB in comparison with the OMP benchmark.
This demonstrates that our proposed algorithms are robust to the perturbation between the mmWave and Sub-6GHz band,
since the CSLW-BOMP and CSLW-OMP algorithm is capable of achieving the balance between the out-of-band spatial information and the greedy OMP and BOMP algorithms
through the carefully designed coefficient $D$ in the recovery of the channel matrix $\mathbf{X}$.

\section{Conclusion} \label{S5}
In this paper, we concentrate on addressing the weak sparsity challenge of near-field channel estimation
resorting to the side information extracted from the spherical wave model and the dual-band communication architecture.
Specifically, the dual-band near-field communication model is elaborated,
where mmWave and Sub-6GHz systems act as co-deployed communication solutions.
The structural characteristic information,
in conjunction with the out-of-band spatial information stemming from the Sub-6GHz band,
is harnessed for enhancing the channel estimation accuracy,
as well as the upper bound of reconstructible sparsity.
A series of variants of the OMP employing side information are proposed,
supported by profound analysis revealing the minimum estimation error.
Numerical simulations are conducted to substantiate the feasibility of the proposed approaches,
indicating fertile advantages in terms of both on- and off-grid estimation accuracy.

\begin{appendices}
    \section{Proof of Theorem 1}
    \begin{IEEEproof}
        \label{Proof1}
        First, we focus on deriving the probability distribution of $|\mathbf{a}_i^{\rm H}\mathbf{y}|^2$.

        Without loss of generality, assume that the first $S$ channel-taps of $\mathbf{x}$ are nonzero,
        i.e., the support set $\mathcal{S}=\{1,2,\cdots,S\}$.
        As a result, $\mathbf{y}$ can be expressed as
        \begin{equation}
            \mathbf{y}=\sum_{j=1}^S\mathbf{a}_jx_j+\mathbf{n},
        \end{equation}
        where $\mathbf{a}_i$ is the $i$-th column of $\mathbf{A}$.
        The correlation term $\mathbf{a}_i^{\rm H}\mathbf{y}$ can thus be
        \begin{equation}
            \mathbf{a}_i^{\rm H}\mathbf{y}=\mathbf{a}_i^{\rm H}\Bigl(\sum_{j=1}^S\mathbf{a}_jx_j+\mathbf{n}\Bigr)=\sum_{i=1}^S(\mathbf{a}_i^{\rm H}\mathbf{a}_j)x_j+\mathbf{a}_i^{\rm H}\mathbf{n}.
        \end{equation}

        Suppose that $M$ is sufficiently large, $\mathbf{a}_i^{\rm H}\mathbf{a}_j$ can be approximated to $M$ when $i=j$ according to the law of large numbers,
        while $\mathbf{a}_i^{\rm H}\mathbf{a}_j \sim \mathcal{CN}(0,M)$ holds for $i \neq j$ by the central limit theorem.
        Similarly, we have $\mathbf{a}_i^{\rm H}\mathbf{n} \sim \mathcal{CN}(0,M\sigma^2)$ \cite{Scarletttsp2013}.
        Consequently, the probability distribution $\mathbf{a}_i^{\rm H}\mathbf{y}$ can be calculated separately for $i \leq S$ and $i > S$.

        For $i \leq S$, there exists $j \in \{1,2,\cdots,S\}$ such that $j=i$, and $\mathbf{a}_i^{\rm H}\mathbf{y}$ can be simplified into
        \begin{equation}
            \begin{aligned}
                \mathbf{a}_i^{\rm H}\mathbf{y} & = \mathbf{a}_i^{\rm H}\mathbf{a}_ix_i+\sum_{j=1,j \neq i}^S(\mathbf{a}_i^{\rm H}\mathbf{a}_j)x_j+\mathbf{a}_i^{\rm H}\mathbf{n} \\
                                               & \approx Mx_i+\sum_{j=1,j \neq i}^S\mathcal{CN}(0,M)x_j+\mathcal{CN}(0,M\sigma^2).
            \end{aligned}
        \end{equation}

        Assume that all nonzero channel-taps $x_i$ have the same modulus $g$ \cite{Scarletttsp2013}, the probability distribution of $\mathbf{a}_i^{\rm H}\mathbf{y}$ satisfies that
        \begin{equation}
            \mathbf{a}_i^{\rm H}\mathbf{y} \sim \mathcal{CN}(Mx_i,M\bigl((S-1)g^2+\sigma^2\bigr)),
        \end{equation}
        which indicates that the correlation term $\mathbf{a}_i^{\rm H}\mathbf{y}$ follows complex Gaussian distribution
        with mean $Mx_i$ and variance $M\bigl((S-1)g^2+\sigma^2\bigr)$.
        Denote $\frac{M}{2}\bigl((S-1)g^2+\sigma^2\bigr)$ as $\sigma_1^2$,
        then $|\mathbf{a}_i^{\rm H}\mathbf{y}|^2$ can be expressed as
        \begin{equation}
            |\mathbf{a}_i^{\rm H}\mathbf{y}|^2=\mathcal{N}(M\Re(x_i),\sigma_1^2)^2+\mathcal{N}(M\Im(x_i),\sigma_1^2)^2,
        \end{equation}
        which follows the non-central $\chi^2$ distribution with degree of freedom $k=2$, noncentrality parameter $a=\sqrt{|\overline{\mathbf{a}_i^{\rm H}\mathbf{y}}|^2}=Mg$ and variance $\sigma_1^2$.

        If $i>S$, for any $j \in \{1,2,\cdots,S\}$, we have $ i \neq j$.
        Therefore, $\mathbf{a}_i^{\rm H}\mathbf{y}$ satisfies that
        \begin{equation}
            \begin{aligned}
                \mathbf{a}_i^{\rm H}\mathbf{y} & = \sum_{i=1}^S\mathcal{CN}(0,M)x_j+\mathcal{CN}(0,M\sigma^2) \\
                                               & \sim \mathcal{CN}(0,M(Sg^2+\sigma^2)).
            \end{aligned}
        \end{equation}

        By letting $\sigma_2^2=\frac{M}{2}(Kg^2+\sigma^2)$,
        the probability distribution of $|\mathbf{a}_i^{\rm H}\mathbf{y}|^2$ for $i=1,2 \cdots N$ is given by
        \begin{equation}
            |\mathbf{a}_i^{\rm H}\mathbf{y}|^2 \sim
            \begin{cases}
                \sigma_1^2\chi'^2(2,Mg) & i \leq S, \\
                \sigma_2^2\chi^2(2)     & i > S,
            \end{cases}
            \label{Corr}
        \end{equation}
        where $\chi^2$ and $\chi'^2$ represents the central and non-central distribution respectively.

        After that, we move on to solving the optimal $v(p_i)$, which is achieved through minimizing the probability of incorrect choices in each iteration of the CLW-OMP algorithm.

        Consider that $i_1 \leq S$ and $i_2>S$, and let $v_1=v(p_{i_1})$, $v_2=v(p_{i_2})$, $T_1=|\mathbf{a}_{i_1}\mathbf{y}|^2$ and $T_2=|\mathbf{a}_{i_2}\mathbf{y}|^2$.
        Then the incorrect index $i_2$ is wrongly chosen instead of correct $i_1$ if and only if $T_1+v_1<T_2+v_2$,
        which is equivalent to $T_1-T_2<v_2-v_1$.
        We denote the probability of this event as $p_e(i_1,i_2)$, and denote $\Delta v=v_2-v_1$ and $T=T_1-T_2$.
        In order to obtain the analytical expression of $v(p_i)$, the following develops the probability distribution function (PDF) of $T$ as a theoretical foundation.

        According to (\ref{Corr}), $T$ is the difference between a non-central $\chi^2$ distribution and a central $\chi^2$ distribution,
        and the characteristic function method can be applied to derive the PDF of $T$.
        For $T_1$ and $T_2$, their characteristic function $\phi(\omega)=\mathbb{E}(e^{j\omega T})$ are given as follows
        \begin{equation}
            \begin{aligned}
                 & \phi_1(\omega)=\frac{1}{1-2j\omega\sigma_1^2}e^{\frac{j\omega M^2g^2}{1-2j\omega \sigma_2^2}}, \\
                 & \phi_2(\omega)=\frac{1}{1-2j\omega\sigma_2^2}.                                                 \\
            \end{aligned}
        \end{equation}

        In addition, since $T_1$ and $T_2$ are independent, the characteristic function of $T$ can be expressed as
        \begin{equation}
            \begin{aligned}
                \phi(\omega) & =\mathbb{E}(e^{j\omega T})=\mathbb{E}(e^{j\omega T_1})\mathbb{E}(e^{-j\omega T_2})                       \\
                             & =\frac{1}{(1-2j\omega\sigma_1^2)(1+2j\omega\sigma_2^2)}e^{\frac{j\omega M^2g^2}{1-2j\omega \sigma_2^2}}.
            \end{aligned}
            \label{Characteristic}
        \end{equation}

        After inverse Fourier transformation, the PDF of $T$ can be obtained based on (\ref{Characteristic}) \cite{Gaussian},
        i.e.,
        \begin{equation}
            p_T(t)=
            \begin{cases}
                \frac{1}{2(\sigma_1^2+\sigma_2^2)}e^{\frac{t}{2\sigma_2^2}}e^{-\frac{M^2g^2}{2(\sigma_1^2+\sigma_2^2)}}                                        & t\geq 0, \\
                \frac{1}{2(\sigma_1^2+\sigma_2^2)}e^{\frac{t}{2\sigma_2^2}}e^{-\frac{M^2g^2}{2(\sigma_1^2+\sigma_2^2)}} \times                                            \\
                Q_1\Bigl(\frac{Mg}{\sigma_1}\sqrt{\frac{\sigma_2^2}{\sigma_1^2+\sigma_2^2}},\sqrt{\frac{t(\sigma_1^2+\sigma_2^2)}{\sigma_1^2\sigma_2^2}}\Bigr) & t<0,
            \end{cases}
            \label{PDF_T}
        \end{equation}
        where $Q_1$ is the first-order Marcum Q-function and given by
        \begin{equation}
            Q_1(a,b)=\int_{b}^{\infty}xe^{-\frac{x^2+a^2}{2}}I_0(ax)dx,
        \end{equation}
        and $I_0(x)$ denotes the modified Bessel function of the first kind with zero order.
        Based on (\ref{PDF_T}), we have
        \begin{equation}
            p_e(i_1,i_2)=P(T<\Delta v).
        \end{equation}

        The expression of $p_e(i_2,i_1)$, which represents the probability of $i_1$ being incorrectly chosen over $i_2$,
        can be similarly calculated as
        \begin{equation}
            p_e(i_2,i_1)=P(T < -\Delta v).
        \end{equation}

        The total error probability $P_e$ in one iteration of CLW-OMP algorithm can be acquired by taking the out-of-band probability $p_{i_1}$ and $p_{i_2}$ into consideration,
        i.e.,
        \begin{equation}
            P_e=p_{i_1}(1-p_{i_2})p_e(i_1,i_2)+p_{i_2}(1-p_{i_1})p_e(i_2,i_1).
        \end{equation}

        Then, $\frac{\partial P_e}{\partial v}=0$ yields the optimal $v(p_i)$.
        However, since $p_T(t)$ is a piecewise function, we need to discuss in cases to further calculate $\frac{\partial P_e}{\partial v}$.

        Consider the situation where $\Delta v > 0$. In this case,
        \begin{equation}
            C=\int_{-\infty}^{0}\frac{1}{2(\sigma_1^2+\sigma_2^2)}e^{\frac{t}{2\sigma_2^2}}e^{-\frac{M^2g^2}{2(\sigma_1^2+\sigma_2^2)}}dt,
            \label{C1}
        \end{equation}
        is a scalar, and $P_e$ can be written as
        \begin{equation}
            \begin{aligned}
                P_e= & p_{i_1}(1-p_{i_2})\Biggl(C+\int_{0}^{\Delta v}\frac{1}{2(\sigma_1^2+\sigma_2^2)}e^{\frac{t}{2\sigma_2^2}}e^{-\frac{M^2g^2}{2(\sigma_1^2+\sigma_2^2)}}            \\
                     & \times Q_1\Biggl(\frac{Mg}{\sigma_1}\sqrt{\frac{\sigma_2^2}{\sigma_1^2+\sigma_2^2}},\sqrt{\frac{t(\sigma_1^2+\sigma_2^2)}{\sigma_1^2\sigma_2^2}}\Biggr)dt\Biggr) \\
                     & + p_{i_2}(1-p_{i_1})\int_{-\infty}^{-\Delta v}\frac{1}{2(\sigma_1^2+\sigma_2^2)}e^{\frac{t}{2\sigma_2^2}}e^{-\frac{M^2g^2}{2(\sigma_1^2+\sigma_2^2)}}dt.
            \end{aligned}
            \label{P_e1}
        \end{equation}

        Combining (\ref{C1}) and (\ref{P_e1}) yields
        \begin{equation}
            \begin{aligned}
                \frac{\partial P_e}{\partial v}= & p_{i_1}(1-p_{i_2})\frac{1}{2(\sigma_1^2+\sigma_2^2)}e^{\frac{\Delta v}{2\sigma_2^2}}e^{-\frac{M^2g^2}{2(\sigma_1^2+\sigma_2^2)}}                               \\
                                                 & \times Q_1\Biggl(\frac{Mg}{\sigma_1}\sqrt{\frac{\sigma_2^2}{\sigma_1^2+\sigma_2^2}},\sqrt{\frac{\Delta v(\sigma_1^2+\sigma_2^2)}{\sigma_1^2\sigma_2^2}}\Biggr) \\
                                                 & -p_{i_2}(1-p_{i_1})\frac{1}{2(\sigma_1^2+\sigma_2^2)}e^{\frac{-\Delta v}{2\sigma_2^2}}e^{-\frac{M^2g^2}{2(\sigma_1^2+\sigma_2^2)}}.
            \end{aligned}
        \end{equation}

        Letting $\frac{\partial P_e}{\partial v}=0$, we have
        \begin{equation}
            e^{\frac{\Delta v}{\sigma_2^2}}Q_1\Biggl(\frac{Mg}{\sigma_1}\sqrt{\frac{\sigma_2^2}{\sigma_1^2+\sigma_2^2}},\sqrt{\frac{\Delta v(\sigma_1^2+\sigma_2^2)}{\sigma_1^2\sigma_2^2}}\Biggr)=\frac{p_{i_2}(1-p_{i_1})}{p_{i_1}(1-p_{i_2})}.
            \label{v-p1}
        \end{equation}

        Since $\Delta v \rightarrow 0$, we obtain that
        \begin{equation}
            Q_1(a,b) \approx \tilde{Q}_1(a,b)=1-\frac{1}{2}(e^{-\frac{a^2-b^2}{2}}-e^{-\frac{a^2+b^2}{2}}),
            \label{Q1est}
        \end{equation}
        for
        \begin{equation}
            \lim\limits_{b \rightarrow 0}(Q_1(a,b)-\tilde{Q}_1(a,b))=0.
        \end{equation}

        By substituting (\ref{Q1est}) into (\ref{v-p1}), the left-hand side of (\ref{v-p1}) can be transformed into
        \begin{equation}
            \begin{aligned}
                 & e^{\frac{\Delta v}{\sigma_2^2}}Q_1\Biggl(\frac{Mg}{\sigma_1}\sqrt{\frac{\sigma_2^2}{\sigma_1^2+\sigma_2^2}},\sqrt{\frac{\Delta v(\sigma_1^2+\sigma_2^2)}{\sigma_1^2\sigma_2^2}}\Biggr)                                                                                     \\
                 & =e^{\frac{\Delta v}{\sigma_2^2}}\Bigl(1-\frac{1}{2}e^{-\frac{M^2g^2\sigma_2^2}{2\sigma_1^2(\sigma_1^2+\sigma_2^2)}}\Bigl(e^{\frac{\Delta v(\sigma_1^2+\sigma_2^2)}{2\sigma_1^2\sigma_2^2}}-e^{-\frac{\Delta v(\sigma_1^2+\sigma_2^2)}{2\sigma_1^2\sigma_2^2}}\Bigr)\Bigr).
            \end{aligned}
        \end{equation}

        Due to Taylor expansion and the fact that $\Delta v \rightarrow 0$,
        we can obtain $e^{\frac{\Delta v}{\sigma_2^2}}\approx1+\frac{\Delta v}{\sigma_2^2}$ and $e^{\frac{\Delta v(\sigma_1^2+\sigma_2^2)}{\sigma_1^2\sigma_2^2}}-e^{-\frac{\Delta v(\sigma_1^2+\sigma_2^2)}{\sigma_1^2\sigma_2^2}} \approx \frac{\Delta v(\sigma_1^2+\sigma_2^2)}{\sigma_1^2\sigma_2^2}$.
        Then,
        \begin{equation}
            \begin{aligned}
                 & e^{\frac{\Delta v}{\sigma_2^2}}\Bigl(1-\frac{1}{2}e^{-\frac{M^2g^2\sigma_2^2}{2\sigma_1^2(\sigma_1^2+\sigma_2^2)}}\Bigl(e^{\frac{\Delta v(\sigma_1^2+\sigma_2^2)}{2\sigma_1^2\sigma_2^2}}-e^{-\frac{\Delta v(\sigma_1^2+\sigma_2^2)}{2\sigma_1^2\sigma_2^2}}\Bigr)\Bigr) \\
                 & =\Bigl(1+\frac{\Delta v}{\sigma_2^2}\Bigr)\Bigl(1-\frac{1}{2}e^{-\frac{M^2g^2\sigma_2^2}{2\sigma_1^2(\sigma_1^2+\sigma_2^2)}} \times \frac{\sigma_1^2+\sigma_2^2}{\sigma_1^2\sigma_2^2}\Delta v\Bigr).
            \end{aligned}
        \end{equation}

        Denoting $A=\frac{1}{\sigma_2^2}$ and $B=\frac{1}{2}e^{\frac{M^2g^2\sigma_2^2}{2\sigma_1^2(\sigma_1^2+\sigma_2^2)}} \times \frac{\sigma_1^2+\sigma_2^2}{\sigma_1^2\sigma_2^2}$,
        we have $A>B$. Therefore, the Marcum Q function term in (\ref{v-p1}) can be simplified into
        \begin{equation}
            \begin{aligned}
                 & e^{\frac{\Delta v}{\sigma_2^2}}Q_1\Biggl(\frac{Mg}{\sigma_1}\sqrt{\frac{\sigma_2^2}{\sigma_1^2+\sigma_2^2}},\sqrt{\frac{\Delta v(\sigma_1^2+\sigma_2^2)}{\sigma_1^2\sigma_2^2}}\Biggr) \\
                 & =(1+A\Delta v)(1-B\Delta v) \approx 1+(A-B)\Delta v.
            \end{aligned}
            \label{v1}
        \end{equation}

        On the other hand, letting $\Delta p=p_{i_2}-p_{i_1} \rightarrow 0$ and $p_{i_1}=p_i$, the right-hand side of (\ref{v-p1}) can be therefore simplified into
        \begin{equation}
            \frac{p_{i_2}(1-p_{i_1})}{p_{i_1}(1-p_{i_2})}=1+\frac{\Delta p}{p_i(1-p_i)}.
            \label{p1}
        \end{equation}

        Combining (\ref{v1}) and (\ref{p1}) yields
        \begin{equation}
            \frac{\Delta v}{\Delta p}=\frac{1}{A-B}\frac{1}{p_i(1-p_i)},
            \label{Derivative1}
        \end{equation}
        which is the derivative of $v(p_i)$ with respect to $p_i$.
        Based on (\ref{Derivative1}), we obtain the optimal expression of the prior factor $v(p_i)$ as follows
        \begin{equation}
            v(p_i)=D\ln\Bigl(\frac{p_i}{1-p_i}\Bigr),
            \label{vp1}
        \end{equation}
        where the coefficient $D$ is given in (\ref{D1}).

        When $\Delta v < 0$, the result in (\ref{vp1}) can be obtained through a similar derivation,
        which completes the proof.
    \end{IEEEproof}

    \section{Proof of Theorem 2}
    \begin{IEEEproof}
        \label{Proof2}
        Based on the proof in Appendix \ref{Proof1}, $\mathbf{A}_i^{\rm H}\mathbf{Y}$ can be expressed as
        \begin{equation}
            \mathbf{A}_i^{\rm H}\mathbf{Y}=\sum_{t=1}^N\mathbf{A}_i^{\rm H}\mathbf{A}_t\mathbf{X}_t+\mathbf{A}_i^{\rm H}\mathbf{N}.
            \label{Correlation2}
        \end{equation}

        Using the assumption that $M$ is sufficiently large, we obtain
        \begin{equation}
            \mathbf{A}_i^{\rm H}\mathbf{A}_t \sim
            \begin{cases}
                \mathcal{CN}(0,M)_{d\times d},                & i \neq t, \\
                \mathcal{M}(M,\mathcal{CN}(0,M))_{d\times d}, & i=t.
            \end{cases}
        \end{equation}

        $\mathcal{CN}(0,M)_{d\times d}$ denotes a $d \times d$ matrix with all its entries i.i.d. as $\mathcal{CN}(0,M)$,
        and $\mathcal{M}(M,\mathcal{CN}(0,M))_{d \times d}$ is defined as
        \begin{equation}
            \begin{aligned}
                 & \mathcal{M}(M,\mathcal{CN}(0,M))_{d \times d}                       \\
                 & =\begin{bmatrix}
                        M                 & \mathcal{CN}(0,M) & \cdots & \mathcal{CN}(0,M) \\
                        \mathcal{CN}(0,M) & M                 & \cdots & \mathcal{CN}(0,M) \\
                        \vdots            & \ddots            & \vdots & \vdots            \\
                        \mathcal{CN}(0,M) & \mathcal{CN}(0,M) & \cdots & M                 \\
                    \end{bmatrix}_{d \times d}.
            \end{aligned}
        \end{equation}

        Note that the entries in $\mathcal{M}(M,\mathcal{CN}(0,M))_{d \times d}$ are also independent.
        Since $\mathbf{A}_i^{\rm H}\mathbf{N}=\mathcal{CN}(0,M\sigma^2)_{d \times K}$,
        $\mathbf{A}_i^{\rm H}\mathbf{Y}$ can be calculated for the cases where $\mathbf{X}_i\neq \mathbf{0}$ and $\mathbf{X}_i=\mathbf{0}$ seperately.

        For $\mathbf{X}_i \neq \mathbf{0}$, there are $S$ nonzero terms in the correlation term in (\ref{Correlation2}),
        and only one of them satisfies that $i=t$.
        With the assumption that all entries in $\mathbf{X}_i$ have the same modulus $g$, $\mathbf{A}_i^{\rm H}\mathbf{Y}$ can be calculated as
        \begin{equation}
            \begin{aligned}
                \mathbf{A}_i^{\rm H}\mathbf{Y}= & \mathcal{M}(M,\mathcal{CN}(0,M))_{d \times d}\mathbf{X}_i                                               \\
                                                & +\sum_{t=1,t\neq i}^{n}\mathcal{CN}(0,M)_{d\times d}\mathbf{X}_t+\mathcal{CN}(0,M\sigma^2)_{d \times K} \\
                =                               & \mathcal{M}(M,\mathcal{CN}(0,M))_{d \times d}\mathbf{X}_i                                               \\
                                                & +\mathcal{CN}(0,M(S-1)dg^2+\sigma^2)_{d \times K}.
            \end{aligned}
        \end{equation}

        Since $\mathcal{M}(M,\mathcal{CN}(0,M))_{d \times d}\mathbf{X}_i=M\mathbf{X}_i+\mathcal{CN}(0,(d-1)Mg^2)_{d\times K}$,
        the expression of $\mathbf{A}_i^{\rm H}\mathbf{Y}$ can be simplified into
        \begin{equation}
            \mathbf{A}_i^{\rm H}\mathbf{Y}=M\mathbf{X}_i+\mathcal{CN}(0,M\bigl((Sd-1)g^2+\sigma^2\bigr))_{d \times K}.
        \end{equation}
        Denoting $M\bigl((Sd-1)g^2+\sigma^2\bigr)$ as $\sigma_1^2$,
        $\|\mathbf{A}_i^{\rm H}\mathbf{Y}\|_F^2$ follows the non-central $\chi^2$ distribution with $2dK$ degrees of freedom,
        and its noncentrality parameter $\lambda$ can be calculated as $\frac{2dKM^2g^2}{\sigma_1^2}$,
        which is similar to the proof in Appendix \ref{Proof1}.
        Compared to the previous proof in Appendix \ref{Proof1}, the difficulty of this theorem lies in the higher degree of freedom of the non-central $\chi^2$ distributions.

        Actually, when the degree of freedom in non-central $\chi^2$ distributions is larger than 2,
        an additional complicated coefficient will appear in its characteristic function,
        which can not be solved by the characteristic function method as the proof given in (\ref{Characteristic}) \cite{Gaussian}.
        As a result, we need to approximate the non-central $\chi^2$ distributed $\|\mathbf{A}_i\mathbf{Y}\|_F^2$ when $\mathbf{X}_i \neq 0$ into a more generalized form.

        According to \cite{Patnaik}, the non-central $\chi^2$ distribution can be approximated to a central $\chi^2$ distribution through Patnaik's second moment approximation, i.e.,
        \begin{equation}
            \chi'^2(n,\lambda) \approx \rho\chi(\tau),
        \end{equation}
        where
        \begin{equation}
            \begin{aligned}
                \rho & =\frac{n+2\lambda}{n+\lambda}=\frac{2dK+\frac{4dKM^2g^2}{\sigma_1^2}}{2dK+\frac{2dKM^2g^2}{\sigma_1^2}},                   \\
                \tau & =\frac{(n+\lambda)^2}{n+2\lambda}=\frac{\Bigl(2dK+\frac{2dKM^2g^2}{\sigma_1^2}\Bigr)^2}{2dK+\frac{4dKM^2g^2}{\sigma_1^2}}.
            \end{aligned}
        \end{equation}
        Recall that central $\chi^2$ distribution is a special case of Gamma distribution, we obtain that
        \begin{equation}
            \begin{aligned}
                \|\mathbf{A}_i^{\rm H}\mathbf{Y}\|_F^2 & =\frac{1}{2}\sigma_1^2\chi'^2(2dK,\lambda)                   \\
                                                       & \approx \frac{1}{2}\sigma_1^2\rho\chi^2(\tau)                \\
                                                       & = \Gamma\Bigl(\frac{\tau}{2},\frac{1}{\rho\sigma_1^2}\Bigr).
            \end{aligned}
        \end{equation}

        When $\mathbf{X}_i=\mathbf{0}$, we can similarly obtain the distribution of $\mathbf{A_i}\mathbf{Y}$ as follows.
        \begin{equation}
            \begin{aligned}
                \mathbf{A}_i^{\rm H}\mathbf{Y} & =\sum_{i=1}^{n}\mathcal{CN}(0,M)_{d \times d}\mathbf{X}_t+\mathcal{CN}(0,M\sigma^2)_{d \times K} \\
                                               & =\sum_{i=1}^n\mathcal{CN}(0,dMg^2)_{d\times K}+\mathcal{CN}(0,M\sigma^2)_{d \times K}            \\
                                               & =\mathcal{CN}(0,M(Sdg^2+\sigma^2))_{d\times K}.
            \end{aligned}
        \end{equation}

        Let $\sigma_2^2=M(Sdg^2+\sigma^2)$.
        As a result, the distribution of $\|\mathbf{A}_i^{\rm H}\mathbf{Y}\|_F^2$ can be expressed as a Gamma distribution in the shape-rate form, i.e.,
        \begin{equation}
            \|\mathbf{A}_i^{\rm H}\mathbf{Y}\|_F^2=\frac{1}{2}\sigma_2^2\chi^2(2dK)=\Gamma\Bigl(dK,\frac{1}{\sigma_2^2}\Bigr).
        \end{equation}

        After obtaining the probability distribution of $\|\mathbf{A}_i^{\rm H}\mathbf{Y}\|_F^2$ in different cases,
        the optimal prior factor $v(p_i)$, which minimizes the error probability, is solved similar to that in Appendix \ref{Proof1}.

        Let $\alpha_1=\frac{\tau}{2}$, $\beta_1=\frac{1}{\rho\sigma_1^2}$ and $\alpha_2=dK$, $\beta_2=\frac{1}{\sigma_2^2}$,
        then $\|\mathbf{A}_i^{\rm H}\mathbf{Y}\|_F^2$ follows the distribution of $\Gamma(\alpha_1,\beta_1)$ and $\Gamma(\alpha_2,\beta_2)$
        when $\mathbf{X}_i=\mathbf{0}$ and $\mathbf{X}_i \neq \mathbf{0}$, respectively.
        Define $v_1$, $v_2$, $T_1$, $T_2$, $\Delta v$, $\Delta p$, $p_i$ as the same in Theorem \ref{Thm31}.
        The PDF of $T=T_1-T_2$ still needs to be derived.

        According to \cite{Gamma}, the PDF of $T$, which is the difference between two Gamma distributions, can be written as
        \begin{equation}
            \begin{aligned}
                p_T(t)=
                \begin{cases}
                    \frac{\tilde{c}}{\Gamma(\alpha_1)}t^{\frac{\alpha_1+\alpha_2}{2}-1}e^{\frac{\beta_2-\beta_1}{2}t}        \\
                    W_{\frac{\alpha_1-\alpha_2}{2},\frac{1-\alpha_1-\alpha_2}{2}}\bigl((\beta_1+\beta_2)t\bigr)  & t \geq 0, \\
                    \frac{\tilde{c}}{\Gamma(\alpha_2)}(-t)^{\frac{\alpha_1+\alpha_2}{2}-1}e^{\frac{\beta_2-\beta_1}{2}t}     \\
                    W_{\frac{\alpha_2-\alpha_1}{2},\frac{1-\alpha_1-\alpha_2}{2}}\bigl(-(\beta_1+\beta_2)t\bigr) & t<0,
                \end{cases}
            \end{aligned}
        \end{equation}
        where $\tilde{c}=\frac{\beta_1^{\alpha_1}\beta_2^{\alpha_2}}{(\beta_1+\beta_2)^{(\alpha_1+\alpha_2)/2}}$ is a scalar,
        $\Gamma(x)$ represents the Gamma function,
        and $W_{\kappa,\mu}(z)$ is the Whittaker function \cite{Whittaker}, which is the solution for the following differential equation
        \begin{equation}
            \frac{d^2w}{dz^2}+\Bigl(-\frac{1}{4}+\frac{\kappa}{z}+\frac{\frac{1}{4}-\mu^2}{z^2}\Bigr)w=0.
        \end{equation}

        Based on Theorem \ref{Thm31}, the derivative of error probability $\frac{\partial P_e}{\partial v}$ when $\Delta v >0$ is given by
        \begin{equation}
            \begin{aligned}
                \frac{\partial P_e}{\partial v}= & p_{i_1}(1-p_{i_2})\frac{\tilde{c}}{\Gamma(\alpha_1)}(\Delta v)^{\frac{\alpha_1+\alpha_2}{2}-1}e^{\frac{\beta_2-\beta_1}{2}\Delta v} \times    \\
                                                 & W_{\frac{\alpha_1-\alpha_2}{2},\frac{\alpha_1+\alpha_2-1}{2}}\bigl((\beta_1+\beta_2)\Delta v\bigr)                                            \\
                                                 & - p_{i_2}(1-p_{i_1})\frac{\tilde{c}}{\Gamma(\alpha_2)}(\Delta v)^{\frac{\alpha_1+\alpha_2}{2}-1}e^{-\frac{\beta_2-\beta_1}{2}\Delta v} \times \\
                                                 & W_{\frac{\alpha_2-\alpha_1}{2},\frac{\alpha_1+\alpha_2-1}{2}}\bigl((\beta_1+\beta_2)\Delta v\bigr).
            \end{aligned}
            \label{Pe'2}
        \end{equation}

        In (\ref{Pe'2}), we use the property of $W_{\kappa,-\mu}(z)=W_{\kappa,\mu}(z)$ for Whittaker functions \cite{Whittaker}.
        From \cite{Whittaker}, the Whittaker function has a concise estimation, i.e.,
        \begin{equation}
            W_{\kappa,\mu}(z)=\frac{\Gamma(2\mu)}{\Gamma(\frac{1}{2}+\mu-\kappa)}z^{\frac{1}{2}-\mu},
            \label{West}
        \end{equation}
        when $\Re(\mu)=\frac{\alpha_1+\alpha_2-1}{2}>\frac{1}{2}$ and $z=(\beta_1+\beta_2)\Delta v \rightarrow 0$,
        which is naturally satisfied in (\ref{Pe'2}) since $\alpha_1,\alpha_2>1$ and $\Delta v \rightarrow 0$.
        By leveraging (\ref{West}) and letting $\frac{\partial P_e}{\partial v}=0$,
        (\ref{Pe'2}) can be finally expressed by
        \begin{equation}
            e^{(\beta_2-\beta_1)\Delta v}=\frac{p_{i_2}(1-p_{i_1})}{p_{i_1}(1-p_{i_2})},
        \end{equation}
        where $\frac{p_{i_2}(1-p_{i_1})}{p_{i_1}(1-p_{i_2})}=1+\frac{\Delta p}{p_i(1-p_i)}$ since $\Delta p \rightarrow 0$,
        which has been demonstrated in Appendix \ref{Proof1}.

        Therefore, since $\Delta p \rightarrow 0$, $\Delta v$ can be expressed as
        \begin{equation}
            \begin{aligned}
                \Delta v & =\frac{1}{\beta_2-\beta_1}\ln(1+\frac{\Delta p}{p_i(1-p_i)})  \\
                         & \approx \frac{1}{\beta_2-\beta_1}\frac{\Delta p}{p_i(1-p_i)}.
            \end{aligned}
            \label{v2}
        \end{equation}

        Based on (\ref{v2}), the derivative of $v(p_i)$ can be obtained,
        which finally gives the expression of $v(p_i)$, i.e.,
        \begin{equation}
            v(p_i)=\frac{1}{\beta_2-\beta_1}\ln\Bigl(\frac{p_i}{1-p_i}\Bigr).
            \label{vp2}
        \end{equation}
        By denoting $\frac{1}{\beta_2-\beta_1}$ as coefficient $D$,
        the optimal form of $v(p_i)$ in (\ref{D2}) is therefore acquired.

        For the case where $\Delta v <0$, (\ref{vp2}) can be obtained by similar deductions.
        This completes the whole proof.

    \end{IEEEproof}
\end{appendices}

\begin{IEEEbiography}
    [{\includegraphics[width=1in,height=1.25in,clip,keepaspectratio]{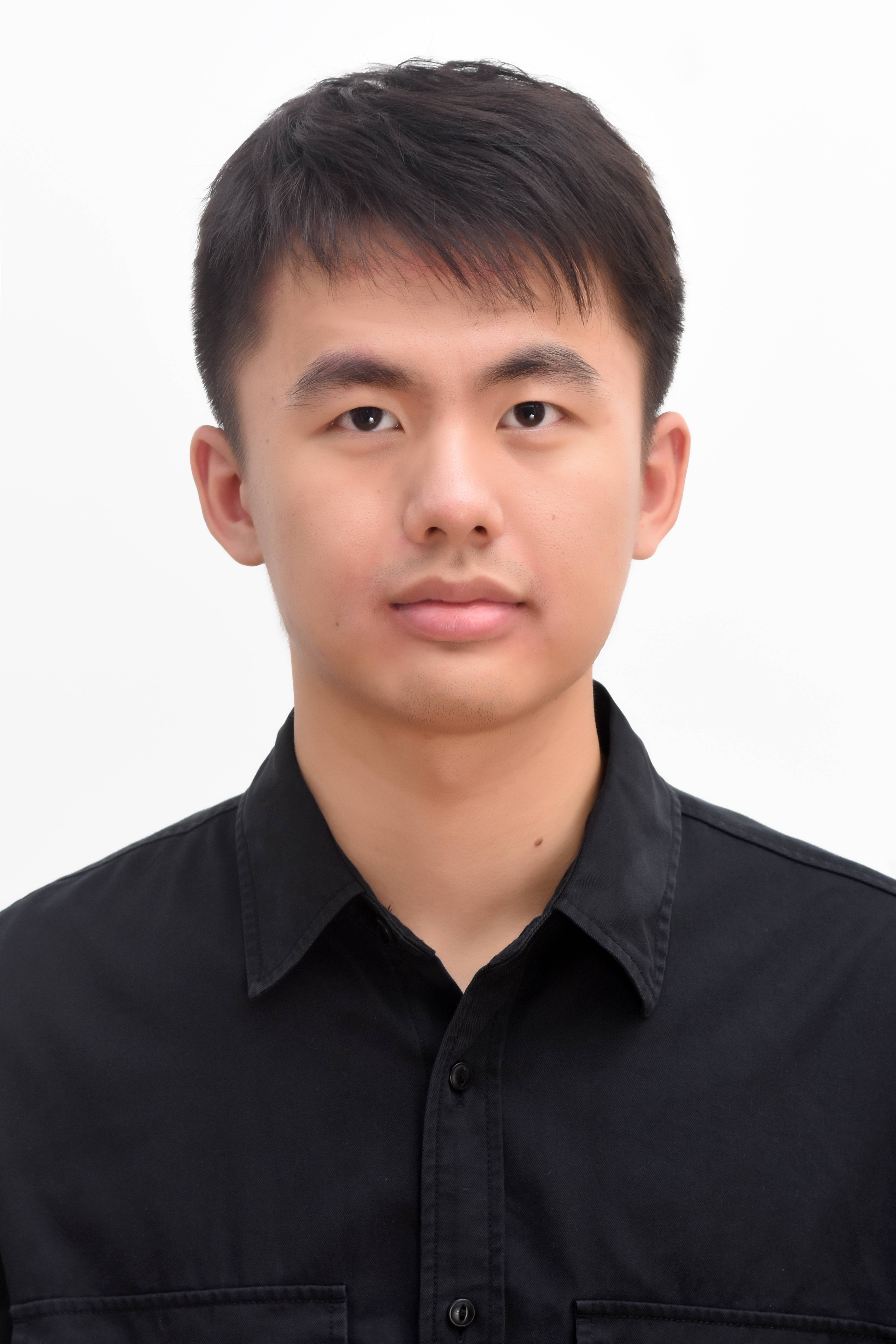}}]
    {Haochen Wu}
    (Student Member, IEEE) received the B.S. (Hons.) degree from the Department of Electronic Engineering, Tsinghua University, Beijing, China, in 2023,
    where he is currently pursuing the Ph.D. degree.
    His main research interests include wireless communications, signal processing, compressive sensing,
    and channel estimation and precoding in near-field communication systems.     
\end{IEEEbiography}

\begin{IEEEbiography}
    [{\includegraphics[width=1in,height=1.25in,clip,keepaspectratio]{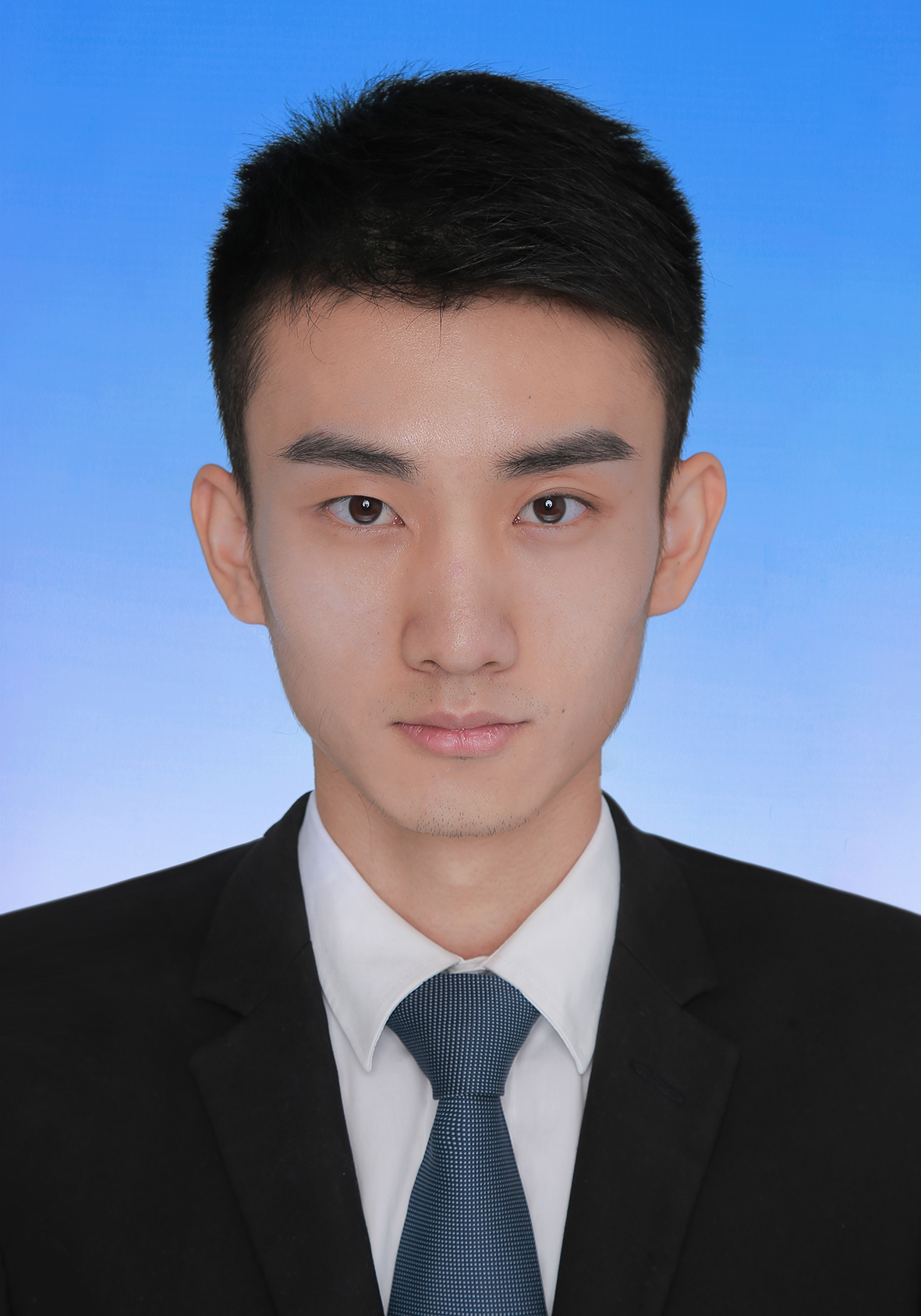}}]
    {Liyang Lu}
    (Member, IEEE) received the B.S. degree from the School of Information Engineering, Beijing University of Posts and Telecommunications (BUPT), China, in 2017, and the Ph.D. degree from the School of Artificial Intelligence, BUPT, in 2022.
    He is currently a postdoctoral fellow at Tsinghua University.
    His area of research interests include compressed sensing, cognitive radios, and MIMO communications.
\end{IEEEbiography}
    
\begin{IEEEbiography}
    [{\includegraphics[width=1in,height=1.25in,clip,keepaspectratio]{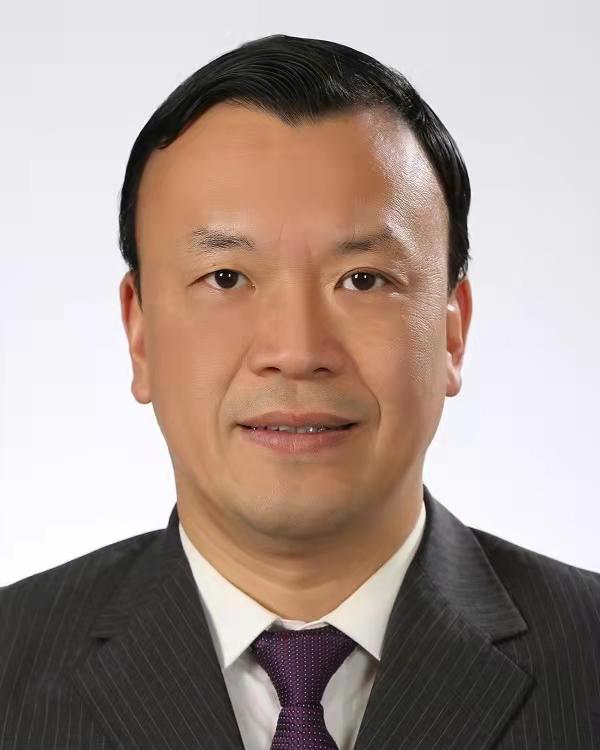}}]
    {Zhaocheng Wang}
    (Fellow, IEEE)
    received the B.S., M.S., and Ph.D. degrees from Tsinghua University in 1991, 1993, and 1996, respectively.

    From 1996 to 1997, he was a Post-Doctoral Fellow with Nanyang Technological University, Singapore.
    From 1997 to 1999, he was a Research Engineer/a Senior Engineer with OKI Techno Centre (Singapore) Pte. Ltd., Singapore. 
    From 1999 to 2009, he was a Senior Engineer/a Principal Engineer with Sony Deutschland GmbH, Germany. 
    Since 2009, he has been a Professor with the Department of Electronic Engineering, Tsinghua University,
    where he is currently the Director of the Broadband Communication Key Laboratory, 
    Beijing National Research Center for Information Science and Technology (BNRist).
    He has authored or coauthored two books, which have been selected by IEEE Press Series on Digital and Mobile Communication (Wiley-IEEE Press).
    He has also authored/coauthored more than 200 peer-reviewed journal articles. 
    He holds 60 U.S./EU granted patents (23 of them as the first inventor).
    His research interests include wireless communications, millimeter wave communications, and optical wireless communications.

    Prof. Wang is a fellow of the Institution of Engineering and Technology.
    He was a recipient of the ICC2013 Best Paper Award, the OECC2015 Best Student Paper Award, the 2016 IEEE Scott Helt Memorial Award,
    the 2016 IET Premium Award, the 2016 National Award for Science and Technology Progress (First Prize), the ICC2017 Best Paper Award,
    the 2018 IEEE ComSoc Asia–Pacific Outstanding Paper Award, and the 2020 IEEE ComSoc Leonard G. Abraham Prize.
    He is currently an Associate Editor of IEEE TRANSACTIONS ON COMMUNICATIONS and IEEE/OSA JOURNAL OF LIGHTWAVE TECHNOLOGY.

\end{IEEEbiography}

\begin{thebibliography}{10}
    \bibitem{wangzhe2024survey}
    Z.~Wang et al., ``A tutorial on extremely large-scale MIMO for 6G: Fundamentals, signal processing, and applications,'' \emph{IEEE Commun. Surveys Tuts.}, to appear, 2024.
    \bibitem{dai2022tcom}
    M.~Cui and L.~Dai, ``Channel estimation for extremely large-scale MIMO: Far-field or near-field?,'' \emph{IEEE Trans. Commun.}, vol.~70, no.~4, pp.~2663--2677, Apr.~2022.
    \bibitem{ziyuansha2021}
    Z.~Sha and Z.~Wang, ``Channel estimation and equalization for terahertz receiver with RF impairments,'' \emph{IEEE J. Sel. Areas Commun.}, vol.~39, no.~6, pp.~1621--1635, Jun. 2021.
    \bibitem{dai2023mag}
    M.~Cui, Z.~Wu, Y.~Lu, X.~Wei and L.~Dai, ``Near-field MIMO communications for 6G: Fundamentals, challenges, potentials, and future directions,'' \emph{IEEE Commun. Mag.}, vol.~61, no.~1, pp.~40--46, Jan. 2023.
    \bibitem{rayleighdistance}
    K.~T.~Selvan and R.~Janaswamy, ``Fraunhofer and Fresnel distances: Unified derivation for aperture antennas,'' \emph{IEEE Antennas Propag. Mag.}, vol.~59, no.~4, pp.~12--15, Aug. 2017.
    \bibitem{swang2022jsac}
    S.~Wang et al., ``A joint hybrid precoding/combining scheme based on equivalent channel for massive MIMO systems,'' \emph{IEEE J. Sel. Areas Commun.}, vol.~40, no.~10, pp.~2882--2893, Oct. 2022.
    \bibitem{zhao2016access}
    Z.~Wang, P.~Zhao, C.~Qian and S.~Chen, ``Location-aware channel estimation enhanced TDD based massive MIMO,'' \emph{IEEE Access}, vol.~4, pp.~7828--7840, Nov. 2016.
    \bibitem{liyang2022}
    L.~Lu, W.~Xu, Y.~Wang, and Z.~Tian, ``Recovery conditions of sparse signals using orthogonal least squares-type algorithms,'' \emph{IEEE Trans. Signal Process.}, vol.~70, pp.~4727--4741, Oct. 2022.
    \bibitem{chenyuanbin2023}
    X.~Guo, Y.~Chen and Y.~Wang, ``Compressed channel estimation for near-field XL-MIMO using triple parametric decomposition," \emph{IEEE Trans. Veh. Technol.}, vol.~72, no.~11, pp.~15040--15045, Nov. 2023.
    \bibitem{ylu2023}
    Y.~Lu and L.~Dai, ``Near-field channel etimation in mixed LoS/NLoS environments for extremely large-scale MIMO systems,'' \emph{IEEE Trans. Commun.}, vol.~71, no.~6, pp.~3694--3707, Jun. 2023.
    \bibitem{xwei2022}
    X.~Wei and L.~Dai, ``Channel estimation for extremely large-scale massive MIMO: Far-field, near-field, or hybrid-field?,'' \emph{IEEE Commun. Lett.}, vol.~26, no.~1, pp.~177--181, Jan. 2022.
    \bibitem{Wyner1976}
    A.~Wyner and J.~Ziv, ``The rate-distortion function for source coding with side information at the decoder,'' \emph{IEEE Trans. Inf. Theory}, vol.~22, no.~1, pp.~1--10, Jan. 1976.
    \bibitem{liyang2023arxiv}
    L.~Lu, Z.~Wang and S.~Chen, ``Joint block-sparse recovery using simultaneous BOMP/BOLS,'' \emph{arXiv:2304.03600}, Apr.~2023.
    \bibitem{outofband2017}
    N.~Gonz\'{a}lez-Prelcic, A.~Ali, V.~Va and R.~W.~Heath, ``Millimeter-wave communication with out-of-band information,'' \emph{IEEE Commun. Mag.}, vol.~55, no.~12, pp.~140--146, Dec. 2017.
    \bibitem{outofband2018}
    A.~Ali, N.~Gonz\'{a}lez-Prelcic and R.~W.~Heath, ``Millimeter wave beam-selection using out-of-band spatial information,'' \emph{IEEE Trans. Wireless Commun.}, vol.~17, no.~2, pp.~1038--1052, Feb. 2018.
    \bibitem{outofbandpdfPeter2016}
    M.~Peter et al. Measurement campaigns and initial channel models for preferred suitable frequency ranges. Accessed: Mar. 2016.
        [Online]. Available: https://bscw.5g-mmmagic.eu/pub/bscw.cgi/d94832/mmMAGIC\_D2-1.pdf
    \bibitem{Scarletttsp2013}
    J.~Scarlett, J.~S.~Evans and S.~Dey, ``Compressed sensing with prior information: Information-theoretic limits and practical decoders,'' \emph{IEEE Trans. Signal Process.}, vol.~61, no.~2, pp.~427--439, Jan. 2013.
    \bibitem{liyang2019wcnc}
    L.~Lu, W.~Xu, Y.~Cui, M.~Dai and J.~Long, ``Block spectrum sensing based on prior information in cognitive radio networks,'' in \emph{Proc. IEEE Wireless Commun. Netw. Conf. (WCNC)}, Marrakesh, Morocco, 2019, pp. 1-5.
    \bibitem{liyang2019}
    L.~Lu, W.~Xu, Y.~Cui, Y.~Dang, S.~Wang, ``Gamma-distribution-based logit weighted block orthogonal matching pursuit for compressed sensing,'' \emph{Electron. Lett.}, vol.~55, no.~17, pp.~959--961, Aug. 2019.
    \bibitem{Angular1}
    D.~Tse and P.~Viswanath, ``Fundamentals of wireless communication,'' \emph{Cambridge, U.K. Cambridge Univ. Press}, 2005.
    \bibitem{Angular2}
    A.~M.~Sayeed, ``Deconstructing multiantenna fading channels,'' \emph{IEEE Trans. Signal Process.}, vol.~50, no.~10, pp.~2563--2579, Oct, 2002.
    \bibitem{Sparse}
    C.~Huang, L.~Liu, C.~Yuen and S.~Sun, ``Iterative channel estimation using LSE and sparse message passing for mmWave MIMO systems,'' \emph{IEEE Trans. Signal Process.}, vol.~67, no.~1, pp.~245--259, Jan., 2019.
    \bibitem{Spherical}
    J.~Sherman, ``Properties of focused apertures in the Fresnel region,'' \emph{IRE Trans. Antennas and Propag.}, vol.~10, no.~4, pp.~399--408, Jul. 1962.
    \bibitem{Block1}
    Y.~C.~Eldar, P.~Kuppinger and H.~Bolcskei, ``Block-sparse signals: Uncertainty relations and efficient recovery,'' \emph{IEEE Trans. Signal Process.}, vol.~58, no.~6, pp.~3042--3054, Jun. 2010.
    \bibitem{Block2}
    A.~Liu, V.~K.~N.~Lau and W.~Dai, ``Exploiting burst-sparsity in massive MIMO with partial channel support information,'' \emph{IEEE Trans. Wireless Commun.}, vol.~15, no.~11, pp.~7820--7830, Nov. 2016.
    \bibitem{Samimi2016TMTT}
    M.~K.~Samimi and T.~S.~Rappaport, ``3-D millimeter-wave statistical channel model for 5G wireless system design,'' \emph{IEEE Trans. Microw. Theory Techn.}, vol.~64, no.~7, pp.~2207--2225, Jul. 2016.
    \bibitem{Kyosti2016EuCAP}
    P.~Kyösti, I.~Carton, A.~Karstensen, W.~Fan and G.~F.~Pedersen, ``Frequency dependency of channel parameters in urban LOS scenario for mmwave communications,'' in \emph{Proc. Eur. Conf. Antennas Propag. (EuCAP)}, Apr. 2016, pp.~1--5.
    \bibitem{LiAccess}
    Z.~Li, C.~Zhang, I.~-T.~Lu and X.~Jia, ``Hybrid precoding using out-of-band spatial information for multi-user multi-RF-chain millimeter wave systems,'' \emph{IEEE Access}, vol.~8, pp.~50872--50883, Mar. 2020.
    \bibitem{XiuAccess}
    Y.~Xiu, W.~Wang and Z.~Zhang, ``A message passing approach to acquire mm-Wave channel state information based on out-of-band data,'' \emph{IEEE Access}, vol.~6, pp.~45665--45680, Feb. 2018.
    \bibitem{MaKeVTC}
    K.~Ma, P.~Zhao and Z.~Wang, ``Deep learning assisted beam prediction using out-of-band information,'' in \emph{Proc. IEEE Veh. Technol. Conf. (VTC)}, Antwerp, Belgium, May 2020, pp.~1--5.
    \bibitem{Support}
    Z.~Gao, C.~Hu, L.~Dai and Z.~Wang, "Channel estimation for millimeter-wave massive MIMO with hybrid precoding over frequency-selective fading channels,'' \emph{IEEE Commun. Lett.}, vol.~20, no.~6, pp.~1259--1262, Jun. 2016.
    \bibitem{MMV}
    H.~Ma, X.~Yuan, L.~Zhou, B.~Li and R.~Qin, ``Joint block support recovery for sub-Nyquist sampling cooperative spectrum sensing,'' \emph{IEEE Wireless Commun. Lett.}, vol.~12, no.~1, pp.~85--88, Jan. 2023.
    \bibitem{Minn2006TCOM}
    Hlaing~Minn, N.~Al--Dhahir and Yinghui~Li, ``Optimal training signals for MIMO OFDM channel estimation in the presence of frequency offset and phase noise,'' \emph{IEEE Trans. Commun.}, vol.~54, no.~10, pp.~1754--1759, Oct. 2006.
    \bibitem{Patnaik}
    B.~Weston, \emph{Approximations to the non central chi-square and noncentral F distributions}.~Texas Tech University, 1978.
    \bibitem{Yuan2023TAP}
    Z.~Yuan, F.~Zhang, Y.~Zhang, J.~Zhang, G.~F.~Pedersen and W.~Fan, ``On phase mode selection in the frequency-invariant beamformer for near-field mmWave channel characterization,'' \emph{IEEE Trans. Antennas Propag.}, vol.~71, no.~11, pp.~8975--8986, Nov. 2023.
    \bibitem{Wang2024TWC}
    K.~Wang et al., ``Knowledge and data dual-driven channel estimation and feedback for ultra-massive MIMO systems under hybrid field beam squint effect,'' \emph{IEEE Trans. Wireless Commun.}, early access, 2024, doi: {http://dx.doi.org/10.1109/TWC.2024.3380638}{10.1109/TWC.2024.3380638}
    \bibitem{Tang2024TCOM}
    A.~Tang et al., ``Line-of-sight extra-large MIMO systems with angular-domain processing: channel representation and transceiver architecture," \emph{IEEE Trans. Commun.}, vol.~72, no.~1, pp.~570--584, Jan. 2024.
    \bibitem{Yang2023CL}
    W.~Yang, M.~Li and Q.~Liu, ``A practical channel estimation strategy for XL-MIMO communication systems,'' \emph{IEEE Commun. Lett.}, vol.~27, no.~6, pp.~1580--1583, Jun. 2023.
    \bibitem{Nayir2022VTC}
    H.~Nayir, E.~Karakoca, A.~Görçin and K.~Qaraqe, ``Hybrid-field channel estimation for massive MIMO systems based on OMP cascaded convolutional autoencoder,'' in \emph{Proc. IEEE 96th Vech. Technol. Conf. (VTC-Fall)}, 2022, pp. 1-6.
    \bibitem{Choi2020TCOM}
    S.~H.~Lim, S.~Kim, B.~Shim and J.~W.~Choi, ``Efficient beam training and sparse channel estimation for millimeter wave communications under mobility,'' \emph{IEEE Trans. Commun.}, vol.~68, no.~10, pp.~6583--6596, Oct. 2020.
    \bibitem{Kim2022ICASSP}
    H.~Chung and S.~Kim, ``Efficient two-stage beam training and channel estimation for RIS-aided mmWave systems via fast alternating least squares,'' in \emph{Proc. IEEE Int. Conf. Acoust. Speech Signal Process. (ICASSP)}, May 2022, pp.~5188--5192.
    \bibitem{Gaussian}
    M.~K.~Simon, ``Probability distributions involving Gaussian random variables: A handbook for engineers and scientists,'' \emph{Berlin, Germany. Springer}, 2006.
    \bibitem{Gamma}
    B.~Klar, ``A note on gamma difference distributions,'' \emph{J. Stat. Comput. Simul.}, vol.~85, no.~18, pp.~3708--3715, 2015.
    \bibitem{Whittaker}
    F.~Olver, D.~Lozier, R.~Boisvert and C.~Clark, \emph{NIST handbook of F distributions}.~Cambridge University Press, 2010.
\end{thebibliography}
\end{document}